\documentclass[
    superscriptaddress,
    reprint,
    floatfix,
    amsmath,
    amssymb,
    aps
]{revtex4-2}

\usepackage{comment}

\providecommand{\plab}[1]{\textbf{#1})}
\providecommand{\pder}[2]{\frac{\partial {#1}}{\partial {#2}}}

\providecommand{\der}[2]{\frac{\dint{#1}}{\dint{#2}}}
\providecommand{\dern}[3]{\frac{{\dint{}^#3} #1}{{\dint{#2}^#3}}}
\providecommand{\pder}[2]{\frac{\partial{#1}}{\partial{#2}}}

\providecommand{\bw}{\begin{widetext}}
        \providecommand{\ew}{\end{widetext}}

\providecommand{\newlinenn}{\\ \nn}

\providecommand{\poissonratio}{\nu_{\scalebox{0.7}{\text{P}}}}

\providecommand{\bzero}{{\bm 0}}

\ExplSyntaxOn

\NewDocumentCommand{\units}{moo}
{
    \ensuremath{

        \IfNoValueTF{#3}{
            \tl_if_blank:nTF{#1}{
                1
            }{
                \clist_use:nn {#1} {\,}
            }
            \IfValueTF{#2}{
                / \int_compare:nNnTF { \clist_count:n {#2} } > { 1 }
                    { ( \clist_use:nn {#2} {\,} ) }
                    { \clist_use:nn {#2} {\,} }
            }{
            }
        }{
            \frac{\clist_use:nn {#1} {\,}}{\clist_use:nn {#2} {\,}}

        }
    }
}

\NewDocumentCommand{\unitsbraces}{moo}
{
    \, [ \units{#1}[#2][#3] ]
}

\NewDocumentCommand{\parameter}{oomoo}
{
    \ensuremath{

        \tl_if_blank:nTF {#1}{

            \IfValueTF{#2}{
                10^{\clist_map_inline:nn {#2} {##1}}
            }{
                1
            }

        }{

            \IfValueTF{#2}{
                \clist_map_inline:nn {#1} {##1}  \times 10^{\clist_map_inline:nn {#2} {##1}}
            }{
                \clist_map_inline:nn {#1} {##1}
            }

        }
        \, \units{#3}[#4][#5]
    }
}
\ExplSyntaxOff

\usepackage{graphicx}  
\usepackage{microtype} 
\usepackage{amsmath}   
\usepackage{hyperref}  

\usepackage{menukeys}
\usepackage{pdfsync}
\usepackage{boldline}
\usepackage[version=3]{acro}
\usepackage{gensymb}
\usepackage{epsfig}
\usepackage{cprotect}
\usepackage{verbatim}
\usepackage{bm}
\usepackage{etoolbox}
\usepackage{multirow}
\usepackage{mathrsfs}
\usepackage{enumitem}
\usepackage{empheq}
\usepackage[font=small,labelfont=bf]{caption}
\usepackage[title,toc,titletoc,page]{appendix}
\usepackage[capitalise]{cleveref}
\usepackage{eurosym}
\usepackage{wrapfig}
\usepackage{scalerel}[2016/12/29]
\usepackage{leftidx}
\usepackage{tikz}
\usepackage{xcolor}
\usepackage{fancyvrb} 

\usetikzlibrary{math} 
\crefname{enumi}{Case}{Cases}

\acsetup{
  make-links ,
  pages / display = first ,
  pages / fill    = {, },
  patch / floats  = false
}

\hypersetup{colorlinks=true,
  linkcolor=blue,
  anchorcolor=orange,
  citecolor=red,
  urlcolor=purple
}

\DeclareAcronym{cnrs}{
        short = {CNRS},
        long = {Centre National de la Recherche Scientifique},
}

\DeclareAcronym{le}{
        short = {L},
        long = {left},
}

\DeclareAcronym{md}{
        short = {MD},
        long = {molecular dynamics},
}

\DeclareAcronym{ri}{
        short = {R},
        long = {right},
}

\DeclareAcronym{mc}{
        short = {MC},
        long = {Monte Carlo},
}

\DeclareAcronym{tirfm}{
        short = {TIRFM},
        long = {total internal reflection fluorescence microscopy},
}

\DeclareAcronym{espci}{
short = {ESPCI},
long = {\'{E}cole Sup\'{e}rieure de Physique et de Chimie Industrielles de la Ville de Paris},
}

\DeclareAcronym{irene}{
short = {IRENE},
long = flu{I}d laye{R} finit{E}-eleme{N}t softwar{E},
}

\DeclareAcronym{fe}{
        short = {FE},
        short-plural-form = {FEs},
        long = finite element,
}

\DeclareAcronym{be}{
        short = {BE},
        short-plural-form = {BEs},
        long = boundary element,
}

\DeclareAcronym{dof}{
        short = {DOF},
        short-plural-form = {DOFs},
        long = degree of freedom,
        long-plural-form =  degrees of freedom
}

\DeclareAcronym{ode}{
        short = {ODE},
        short-plural-form = {ODEs},
        long = ordinary differential equation,
        long-plural-form =  ordinary differential equations,
}

\DeclareAcronym{cg}{
        short = {CG},
        long = continuous Galerkin,
}

\DeclareAcronym{ref}{
        short = \reference,
        short-plural-form = {{\reference}s},
        long = reference,
}

\DeclareAcronym{cur}{
        short = \current,
        short-plural-form = {{\current}s},
        long = current,
}

\DeclareAcronym{ale}{
        short = {ALE},
        short-plural-form = {ALEs},
        long = arbitrary Lagrangian-Eulerian ,
}

\DeclareAcronym{fl}{
        short = {F},
        short-plural-form = {Fs},
        long = fluid,
}

\DeclareAcronym{al}{
        short = {AL},
        short-plural-form = {ALs},
        long = arc length ,
        long-plural-form = arc lengths,
}

\DeclareAcronym{tmp}{
        short = {TMP},
        short-plural-form = {TMPs},
        long = {trans-membrane protein},
        long-plural-form =  {trans-membrane proteins},
}

\DeclareAcronym{lb}{
        short = {LB},
        long = \ac{lb},
}

\DeclareAcronym{vp}{
        short = {VP},
        short-plural-form = {VPs},
        long = variational problem,
        long-plural-form = variational problems,
}

\DeclareAcronym{cn}{
        short = {CN},
        long = Crank-Nicolson,
}

\DeclareAcronym{fsi}{
        short = {FSI},
        long = fluid-structure interaction,
        long-plural-form = fluid-structure interactions,
}

\DeclareAcronym{ipcs}{
        short = {IPCS},
        long = incremental pressure correction scheme,
}

\DeclareAcronym{lhs}{
        short = {LHS},
        short-plural-form = {LHSs},
        long = left-hand side,
        long-plural-form = left-hand sides,
}
\DeclareAcronym{rhs}{
        short = {RHS},
        short-plural-form = {RHSs},
        long = right-hand side,
        long-plural-form = right-hand sides,
}

\DeclareAcronym{bc}{
        short = {BC},
        short-plural-form = {BCs},
        long = {boundary condition},
        long-plural-form = {boundary conditions}
}

\DeclareAcronym{bvp}{
        short = {BVP},
        short-plural-form = {BVPs},
        long = {boundary-value problem},
        long-plural-form = {boundary-value problems}
}

\DeclareAcronym{ns}{
        short = {NS},
        long = {Navier-Stokes},
}

\DeclareAcronym{fem}{
        short = {FEM},
        short-plural-form = {FEMs},
        long = {\ac{fe} method},
        long-plural-form = {finite-element methods}
}

\DeclareAcronym{pde}{
        short = {PDE},
        short-plural-form = {PDEs},
        long = {partial differential equation},
        long-plural-form = {partial differential equations}
}
\DeclareAcronym{fenics}{
        short = {FEniCS},
        long = {Finite Element Computational Software},
}

\definecolor{glossary-link-color}{RGB}{129, 19, 49}
\definecolor{revision-color}{RGB}{100,0,0}

\providecommand{\mlineeq}{{\mathord{%
                \tikz[baseline=-0.1ex]{
                    \tikzmath{\size = 0.06ex;  }
                    \draw[line width=1pt, draw opacity =1] (0,0) -- (\size,0);
                }%
            }}}
\providecommand{\mlineineq}{{\mathord{%
                \tikz[baseline=-0.1ex]{

                    \tikzmath{\size = 0.06ex;  }
                    \draw[line width=1pt, draw opacity =0.5] (0,0) -- (\size,0);
                    \filldraw[color=black, fill=black](0,0) circle (0.05);
                }
            }}}

\providecommand{\mlineouteq}{{\mathord{%
                \tikz[baseline=-0.1ex]{

                    \tikzmath{\size = 0.06ex;  }
                    \draw[line width=1pt, draw opacity =0.5] (0,0) -- (\size,0);
                    \filldraw[color=black, fill=black](\size,0) circle (0.05);
                }
            }}}

\providecommand{\mwalleq}{{\mathord{%
                \tikz[baseline=-0.1ex]{

                    \tikzmath{\size = 0.03ex;  }

                    \draw[line width=1pt] (0,0) -- (\size,0);
                    \draw[line width=1pt, draw opacity = 0.15] (\size,0) -- (\size,\size);
                    \draw[line width=1pt] (\size,\size) -- (0,\size);
                    \draw[line width=1pt, draw opacity = 0.15] (0,\size) -- (0,0);
                }%
            }}}

\providecommand{\mtopeq}{{\mathord{%
                \tikz[baseline=-0.1ex]{

                    \tikzmath{\size = 0.03ex;  }

                    \draw[line width=1pt, draw opacity = 0.15] (0,0) -- (\size,0);
                    \draw[line width=1pt, draw opacity = 0.15] (\size,0) -- (\size,\size);
                    \draw[line width=1pt] (\size,\size) -- (0,\size);
                    \draw[line width=1pt, draw opacity = 0.15] (0,\size) -- (0,0);
                }%
            }}}
\providecommand{\mbottomeq}{{\mathord{%
                \tikz[baseline=-0.1ex]{

                    \tikzmath{\size = 0.03ex;  }

                    \draw[line width=1pt] (0,0) -- (\size,0);
                    \draw[line width=1pt, draw opacity = 0.15] (\size,0) -- (\size,\size);
                    \draw[line width=1pt, draw opacity = 0.15] (\size,\size) -- (0,\size);
                    \draw[line width=1pt, draw opacity = 0.15] (0,\size) -- (0,0);
                }%
            }}}

\providecommand{\mleftbottomeq}{{\mathord{%
                \tikz[baseline=-0.1ex]{

                    \tikzmath{\size = 0.03ex;  }

                    \draw[line width=1pt, draw opacity = 0.15] (0,0) -- (\size,0);
                    \draw[line width=1pt, draw opacity = 0.15] (\size,0) -- (\size,\size);
                    \draw[line width=1pt, draw opacity = 0.15] (\size,\size) -- (0,\size);
                    \draw[line width=1pt, draw opacity = 0.15] (0,\size) -- (0,0);

                    \filldraw[color=black, fill=black](0,0) circle (0.04);

                }%
            }}}

\providecommand{\myineq}{{\mathord{%
                \tikz[baseline=-0.1ex]{

                    \tikzmath{\size = 0.03ex;  }

                    \draw[line width=1pt, draw opacity = 0.15] (0,0) -- (\size,0);
                    \draw[line width=1pt, draw opacity = 0.15] (\size,0) -- (\size,\size);
                    \draw[line width=1pt, draw opacity = 0.15] (\size,\size) -- (0,\size);
                    \draw[line width=1pt] (0,\size) -- (0,0);
                }%
            }}}

\providecommand{\myouteq}{{\mathord{%
                \tikz[baseline=-0.1ex]{

                    \tikzmath{\size = 0.03ex;  }

                    \draw[line width=1pt, draw opacity = 0.15] (0,0) -- (\size,0);
                    \draw[line width=1pt] (\size,0) -- (\size,\size);
                    \draw[line width=1pt, draw opacity = 0.15] (\size,\size) -- (0,\size);
                    \draw[line width=1pt, draw opacity = 0.15] (0,\size) -- (0,0);
                }%
            }}}

\providecommand{\msquareeq}{{\mathord{%
                \tikz[baseline=-0.1ex]{

                    \tikzmath{\size = 0.03ex;  }

                    \draw[line width=1pt] (0,0) -- (\size,0);
                    \draw[line width=1pt] (\size,0) -- (\size,\size);
                    \draw[line width=1pt] (\size,\size) -- (0ex,\size);
                    \draw[line width=1pt] (0ex,\size) -- (0ex,0ex);
                }%
            }}}

\providecommand{\mellipseeq}{{\mathord{%
                \tikz[baseline=-0.3ex]{
                    \draw[line width=0.8pt] (0,0) ellipse [x radius=0.6ex, y radius=0.3ex];
                }%
            }}}

\providecommand{\mcircineq}{{\mathord{%
                \tikz[baseline=-0.5ex]{
                    \draw[line width=1pt] (0,0) circle [radius=0.3ex];
                    \draw[line width=1pt, draw opacity = 0.15] (0,0) circle [radius=.7ex];
                }%
            }}}

\providecommand{\mcircouteq}{{\mathord{%
                \tikz[baseline=-0.5ex]{
                    \draw[line width=1pt, draw opacity = 0.15] (0,0) circle [radius=0.3ex];
                    \draw[line width=1pt] (0,0) circle [radius=.65ex];
                }%
            }}}

\providecommand{\libtickzpicture}{
    \begin{tikzpicture}[scale=1, baseline={(current bounding box.center)}]
        \def\s{80}
        \def\A{0.1}
        \def\xmax{1.5}
        \def\scalefactorletters{1.5}
        \def\radtodeg{360.0/(2.0*pi)}
        \def\degtorad{1.0/(360.0/(2.0*pi))}

        \draw[color={rgb,255:red,0; green,0; blue,255}, opacity=0.8, line width=2pt] plot[domain=-0.2:1.2, samples=100]
        (\x*\xmax, {\A*sin(2*pi*\x*\xmax*\radtodeg)});

        \node at (0*\xmax, {\A*sin(2*pi* 0*\xmax * \radtodeg)})
        [rotate= atan(\A *2*pi  * cos(2*pi * 0*\xmax * \radtodeg))] {\scalebox{\scalefactorletters}{i}};
        \node at (0.25*\xmax, {\A*sin(2*pi* 0.25*\xmax * \radtodeg)})
        [rotate= atan(\A *2*pi  * cos(2*pi * 0.25*\xmax * \radtodeg))] {\scalebox{\scalefactorletters}{r}};
        \node at (0.5*\xmax, {\A*sin(2*pi* 0.5*\xmax * \radtodeg)})
        [rotate= atan(\A *2*pi  * cos(2*pi * 0.5*\xmax * \radtodeg))] {\scalebox{\scalefactorletters}{e}};
        \node at (0.75*\xmax, {\A*sin(2*pi* 0.75*\xmax * \radtodeg)})
        [rotate= atan(\A *2*pi  * cos(2*pi * 0.75*\xmax * \radtodeg))] {\scalebox{\scalefactorletters}{n}};
        \node at (1.0*\xmax, {\A*sin(2*pi* 1.0*\xmax * \radtodeg)})
        [rotate= atan(\A *2*pi  * cos(2*pi * 1.00*\xmax * \radtodeg))] {\scalebox{\scalefactorletters}{e}};
    \end{tikzpicture}%
}

\newsavebox{\libnamebox}

\AtBeginDocument{%
    \sbox{\libnamebox}{%
        \raisebox{1.0ex}{%
            \scalebox{0.8}{\libtickzpicture}
        }%
    }%
}

\providecommand{\be}{\begin{equation}}
        \providecommand{\ee}{\end{equation}}
\providecommand{\bsp}{\begin{split}}
        \providecommand{\esp}{\end{split}}
\providecommand{\bea}{\begin{eqnarray}}
        \providecommand{\eea}{\end{eqnarray}}
\providecommand{\beas}{\begin{eqnarray*}}
        \providecommand{\eeas}{\end{eqnarray*}}

\providecommand{\myineqcap}{{\protect \myineq}}

\providecommand{\mellipseeqcap}{{\protect \mellipseeq}}

\providecommand{\mtopeqcap}{{\protect \mtopeq}}
\providecommand{\mbottomeqcap}{{\protect \mbottomeq}}
\providecommand{\mleftbottomeqcap}{{\protect \mleftbottomeq}}

\providecommand{\vbar}{\overline{v}}
\providecommand{\bvbar}{{\bm \vbar}}
\providecommand{\wbar}{\overline{w}}

\providecommand{\evec}[2]{{e_{#1}^{#2}}}
\providecommand{\bevec}[1]{{{\bm e}_{#1}}}
\providecommand{\titleditem}[2]{\item \textbf{#1.} \protect#2}
\providecommand{\mcircineqcap}{{\protect \mcircineq}}
\providecommand{\mcircouteqcap}{{\protect \mcircouteq}}
\providecommand{\om}{\Omega}
\providecommand{\omc}{\Omega^\current}
\providecommand{\omr}{\Omega^\reference}
\providecommand{\pom}{{\partial \om}}
\providecommand{\pom}{{\partial \om}}

\providecommand{\pomineq}{\pom_\myineqcap}
\providecommand{\pominc}{$\pomineqc$}

\providecommand{\pomsqr}{$\pomsqeqr$}
\providecommand{\current}{\textrm{c}}
\providecommand{\reference}{\textrm{r}}

\providecommand{\pomsqeqr}{\pom_\msquareeq^{\reference}}
\providecommand{\pomineqc}{\pom_\myineqcap^{\current}}
\providecommand{\pomineqr}{\pom_\myineqcap^{\reference}}
\providecommand{\pomin}{$\pomineq$}
\providecommand{\pomout}{$\pomouteq$}
\providecommand{\pomouteq}{\pom_\myouteqcap}
\providecommand{\pomouteqc}{\pom_\myouteqcap^{\current}}
\providecommand{\pomouteqr}{\pom_\myouteqcap^{\reference}}

\providecommand{\pomellipseeqc}{{\pom_\mellipseeqcap^{\current}}}
\providecommand{\pomellipsec}{$\pomellipseeqc\;$}
\providecommand{\pomellipser}{$\pomellipseeqr\;$}
\providecommand{\pomellipseeqr}{{\pom_\mellipseeqcap^{\reference}}}

\providecommand{\pomoutc}{$\pomouteqc\;$}
\providecommand{\nmhalf}{{n-1/2}}
\providecommand{\nmthalf}{{n-3/2}}
\providecommand{\nmthreehalf}{{n-3/2}}

\providecommand{\fnij}[4]{{#1}^{#2,\,#3 }_{\phantom{#2,\,} #4}}

\providecommand{\pomcircineq}{\pom_\mcircineqcap}
\providecommand{\pomweq}{\pom_\mwalleqcap}
\providecommand{\pomlineeq}{\pom_\mlineeqcap}
\providecommand{\omlineeq}{\om_\mlineeqcap}
\providecommand{\omlineeqr}{\om_\mlineeqcap^\reference}

\providecommand{\pomline}{$\pom_\mlineeqcap$}
\providecommand{\pomlineineq}{\pom_\mlineineqcap}

\providecommand{\pomlineouteq}{\pom_\mlineouteqcap}
\providecommand{\pomlineout}{$\pom_\mlineouteqcap$}

\providecommand{\mwalleqcap}{{\protect \mwalleq}}
\providecommand{\mlineeqcap}{{\protect \mlineeq}}
\providecommand{\mlineineqcap}{{\protect \mlineineq}}
\providecommand{\mlineouteqcap}{{\protect \mlineouteq}}

\providecommand{\pomlineeqr}{\pomlineeq^\reference}

\providecommand{\pomlineineqr}{\pomlineineq^\reference}
\providecommand{\pomcircineqr}{\pomcircineq^\reference}
\providecommand{\pomlineouteqr}{\pomlineouteq^\reference}

\providecommand{\pomcircouteqc}{\pomcircouteq^\current}
\providecommand{\pomcircouteq}{\pom_\mcircouteqcap}
\providecommand{\pomcircouteqc}{\pomcircouteq^\current}
\providecommand{\pomcircouteqr}{\pomcircouteq^\reference}
\providecommand{\pomweqc}{\pomweq^{\current}}
\providecommand{\pomweqr}{\pomweq^{\reference}}
\providecommand{\pomwc}{$\pomweqc$}

\providecommand{\pomtopeq}{\pom_\mtopeqcap}

\providecommand{\pomtopeqc}{\pom_\mtopeqcap^{\current}}
\providecommand{\pomtopc}{$\pomtopeqc$}
\providecommand{\pombottomc}{$\pombottomeqc$}

\providecommand{\pomtopeqr}{\pom_\mtopeqcap^{\reference}}
\providecommand{\pombottomeq}{\pom_\mbottomeqcap}
\providecommand{\pombottomeqc}{\pom_\mbottomeqcap^{\current}}
\providecommand{\pomleftbottomeqc}{\pom_\mleftbottomeqcap^{\current}}
\providecommand{\pombottomeqr}{\pom_\mbottomeqcap^{\reference}}
\providecommand{\pomleftbottomeqr}{\pom_\mleftbottomeqcap^{\reference}}
\providecommand{\pombottom}{$\pombottomeq$}

\providecommand{\lr}{l_{\reference}}
\providecommand{\lc}{l_{\current}}

\providecommand{\fluid}{\mathrm{\scalebox{0.6}{F}}}
\providecommand{\bod}{\mathrm{\scalebox{0.6}{B}}}
\providecommand{\ela}{\mathrm{\scalebox{0.6}{E}}}
\providecommand{\mem}{\mathrm{\scalebox{0.6}{M}}}
\providecommand{\dom}{\mathrm{\scalebox{0.6}{D}}}

\providecommand{\lapl}{\mathrm{\scalebox{0.6}{L}}}

\providecommand{\pomtop}{$\pomtopeq$}

\providecommand{\kap}{\kappa}
\providecommand{\nab}{\nabla}

\providecommand{\pas}{\text{Pa}}
\providecommand{\met}{\text{m}}

\providecommand{\kg}{\text{kg}}

\providecommand{\second}{\text{s}}

\providecommand{\kel}{K}

\providecommand{\deltat}{{\Delta t}}

\providecommand{\odeltat}{{{\cal O}(\deltat)}}

\providecommand{\myouteqcap}{{\protect \myouteq}}

\providecommand{\baligned}{\begin{equation}\begin{aligned}}
            \providecommand{\ealigned}{\end{aligned}\end{equation}}
\providecommand{\nn}{\nonumber}

\providecommand{\dint}[1]{\text{d} {#1}}

\ProvideDocumentCommand{\crefs}{m}{%
    \begingroup
    \crefname{equation}{}{}%
    \cref{#1}%
    \endgroup
}
\providecommand{\crefineq}[1]{\text{\cref{#1}}}

\providecommand{\xcoord}{x}
\providecommand{\xcartesian}{{\mathsf{x}}}
\newcommand{\xc}[1][]{{\xcartesian_{\current\ifx\empty#1\empty\else\, #1\fi}}}
\newcommand{\xr}[1][]{{\xcartesian_{\reference\ifx\empty#1\empty\else\, #1\fi}}}
\providecommand{\vf}{{v_\fluid}}
\providecommand{\vmem}[1]{v_\mem^{#1}}
\providecommand{\wmem}{w_\mem}
\providecommand{\fmemt}{f_{\mem\, \text{t}}}

\providecommand{\bVf}{{\bm \Vf}}
\providecommand{\bvbarmem}{{\bm \vbarmem}}
\providecommand{\vbarmem}{\vbar_\mem}
\providecommand{\wbarmem}{\wbar_\mem}
\providecommand{\Vmem}{V_\mem}
\providecommand{\phimem}{{\phi_\mem}}
\providecommand{\phif}{{\phi_\fluid}}
\providecommand{\Vf}{V_\fluid}

\providecommand{\vfc}{\vf_\current}
\providecommand{\bvfc}[1]{{\bm{v}}_{\fluid \, \current}^{#1}}
\providecommand{\bvfr}[1]{{\bm{v}_{\fluid \,\reference}^{#1}}}
\providecommand{\vfr}[1]{v_{\fluid \, \reference}^{#1}}
\providecommand{\rhob}{\rho_\bod}
\providecommand{\rhof}{\rho_\fluid}
\providecommand{\rhoela}{\rho_\ela}
\providecommand{\rhom}{\rho_\mem}
\providecommand{\Sela}{{S_\ela}}
\providecommand{\Sdom}{{S_\dom}}
\providecommand{\etaf}{\eta_\fluid}
\providecommand{\etam}{{\eta_\mem}}
\providecommand{\surfacetensionsymbol}{\varsigma}

\providecommand{\sigmaf}{{\surfacetensionsymbol_\fluid}}

\providecommand{\sigmam}[1]{{\surfacetensionsymbol_\mem^{#1}}}
\providecommand{\sigmamast}{{\surfacetensionsymbol_\mem^\ast}}
\providecommand{\sigmafc}[1]{\surfacetensionsymbol_{\fluid \, \current}^{#1}}
\providecommand{\sigmafr}[1]{{\surfacetensionsymbol_{\fluid\, \reference}^{#1}}}
\providecommand{\udom}{u_\dom}
\providecommand{\udomdot}{{\dot{u}_\dom}}
\providecommand{\bu}{{\bm u}}
\providecommand{\budom}{\bu_\dom}
\providecommand{\uela}{u_\ela}
\providecommand{\ueladot}{\dot{u}_\ela}
\providecommand{\buela}{\bu_\ela}
\providecommand{\budomdot}{{\dot{\bu}_\dom}}
\providecommand{\bueladot}{{\dot{\bu}_\ela}}
\providecommand{\umem}{u_\mem}

\providecommand{\bumem}{\bu_\mem}
\providecommand{\bumemdot}{{\dot{\bu}_\mem}}
\providecommand{\bVmem}{{\bm V_\mem}}
\providecommand{\bvmem}[1]{{{\bm v}_\mem^{#1}}}
\providecommand{\Wmem}{W_\mem}
\providecommand{\xmemc}{{\xcartesian_\current^\mem}}
\providecommand{\xmemr}[1]{{\xcartesian_\reference^{\mem\, #1}}}
\providecommand{\bxmemc}{{\bm{\xmemc{}}}}
\providecommand{\bxmemr}{{\bm{\xmemr{}}}}
\providecommand{\bxc}{{\bm{\xcartesian}_{\current}}}
\providecommand{\bxr}{{\bm{\xcartesian}_{\reference}}}
\providecommand{\bnormalr}{\hat{\bm n}^\reference}
\providecommand{\normalmem}{{n_\mem}}
\providecommand{\bnormalc}{\hat{\bm n}^\current}
\providecommand{\normalr}[1]{\hat{n}^{\reference\, #1}}
\providecommand{\normalc}[1]{\hat{n}^{\current\, #1}}

\providecommand{\pderc}[2]{\pder{#1}{\xc[#2]}}
\providecommand{\pderr}[2]{\pder{#1}{\xr[#2]}}

\providecommand{\sigmastressr}[2]{\sigma_{\fluid\, \reference\, #1#2}}
\providecommand{\bsigmastressr}{{\bm \sigma}_{\fluid\, \reference}}
\providecommand{\sigmastressc}[2]{\sigma_{\fluid\, \current\, #1#2}}
\providecommand{\bsigmastressc}{{\bm \sigma}_{\fluid\, \current}}

\providecommand{\structure}{\mathrm{\scalebox{0.6}{S}}}

\providecommand{\madded}{m_{\scalebox{0.6}{\text{A}}}}

\providecommand{\mstr}{m_\structure}

\usepackage{acro}
\usepackage{xparse}
\usepackage[nolong]{glossaries}

\usepackage{setspace}

\DeclareAcronym{feat2d}{
        short = {FEAT2D},
        long = Finite Element Analysis Tools 2D,
}

\DeclareAcronym{bod}{
        short = {B},
        long = rigid body,
        long-plural-form = rigid bodies
}

\DeclareAcronym{ela}{
        short = {E},
        long = elastic body,
        long-plural-form = elastic bodies
}

\DeclareAcronym{flu}{
        short = {F},
        long = fluid,
        long-plural-form = fluids
}

\DeclareAcronym{dom}{
        short = {D},
        long = fluid domain
}

\DeclareAcronym{mem}{
        short = {M},
        long = Helfrich membrane
}

\providecommand{\mtitle}{Interaction between fluids and Helfrich membranes}
\providecommand{\vfbottom}{\bm v_{\fluid \, \mbottomeqcap}}
\providecommand{\vfin}{\bm v_{\fluid \, \myineqcap}}
\providecommand{\bvbarf}{\bvbar_\fluid}
\providecommand{\vbarf}{\vbar_\fluid}
\providecommand{\kaptwod}{\kap_{2\text{d}}}
\providecommand{\sigmafleftbottomc}{{\sigmaf_{\mleftbottomeqcap}}}
\providecommand{\phifleftbottomc}{{\phif_{\mleftbottomeqcap}}}

\providecommand{\bxi}[1]{{\bm \xi}(#1)}
\providecommand{\fflun}{{f_{\fluid\, \text{n}}}  }

\providecommand{\fetamemn}{{f^\eta_{\mem\, \text{n}}}}

\providecommand{\eqsfluidbodyref}{eq_fl_rigid_fl_1_reference,eq_fl_rigid_fl_2_reference,eq_fl_rigid_rigid_reference,bc_fl_rigid_1_reference,bc_fl_rigid_2_reference,bc_fl_rigid_3_reference,bc_fl_rigid_4_reference,bc_fl_rigid_5_reference,bc_fl_rigid_6_reference}
\providecommand{\eqsmembrane}{{eq-dyn-continuity,eq-dyn-v,eq-dyn-w,eq-dyn-dX,eq-dyn-dX1,eq-dyn-dX2}}
\providecommand{\eqsdom}{{eq_mesh_fluid_rigid,eq_mesh_fluid_rigid_dot}}
\providecommand{\eqsbfluc}{{eq_fl_rigid_fl_1_current,eq_fl_rigid_fl_2_current}}

\providecommand{\bcsmembrane}{{bc_fl_mem_mem_1,bc_fl_mem_mem_2,bc_fl_mem_mem_3,bc_fl_mem_mem_4,bc_fl_mem_mem_5,bc_fl_mem_mem_6,bc_fl_mem_mem_7,bc_fl_mem_mem_8}}
\providecommand{\bcsdom}{{bc_fl_mem_dom_1,bc_fl_mem_dom_2,bc_fl_mem_dom_3,bc_fl_mem_dom_4,bc_fl_mem_dom_5,bc_fl_mem_dom_6,bc_fl_mem_dom_7,bc_fl_mem_dom_8}}
\providecommand{\bcbflu}{{bc_fl_membrane_1_current,bc_fl_membrane_2_current,bc_fl_membrane_3_current,bc_fl_membrane_4_current,bc_fl_membrane_5_current,bc_fl_membrane_6_current,bc_fl_membrane_7_current}}

\definecolor{black_color}{RGB}{0,0,0}
\definecolor{acronym_link_color}{RGB}{0,0,100}

\providecommand{\focalpoint}{f}
\providecommand{\bfocalpoint}{{\bm \focalpoint}}
\providecommand{\kel}{K}
\providecommand{\muel}{\mu}

\providecommand{\tbody}{\tau_{\bod}}

\makeatletter
\DeclareRobustCommand{\currentsec}{%
        \ifnum\value{subsubsection}>0
                \Roman{section}.\Alph{subsection}.\arabic{subsubsection}%
        \else\ifnum\value{subsection}>0
                        \Roman{section}.\Alph{subsection}%
                \else
                        \Roman{section}%
                \fi\fi%
}
\makeatother

\acsetup{
    make-links = true,
    format/short = \color{acronym_link_color},    
    format/long = \color{acronym_link_color},     
    list/template=description,
    pages/display=none
}
\allowdisplaybreaks

\crefname{paragraph}{Section}{paragraphs}
\Crefname{paragraph}{Sections}{Paragraphs}

\date{}

\begin{document}

\setlength{\parskip}{5pt plus 0pt minus 0pt}

\title{\mtitle}

\author{Dennis W\"{o}rthm\"{u}ller}
\affiliation{Institut Curie, PSL Research University, Paris, France}
\affiliation{CNRS UMR168, 11 rue Pierre et Marie Curie, 75005, Paris, France}

\author{Pierre Sens}
\affiliation{Institut Curie, PSL Research University, Paris, France}
\affiliation{CNRS UMR168, 11 rue Pierre et Marie Curie, 75005, Paris, France}

\author{Michele Castellana}
\thanks{Corresponding author: \href{mailto:michele.castellana@curie.fr}{michele.castellana@curie.fr}}
\affiliation{Institut Curie, PSL Research University, Paris, France}
\affiliation{CNRS UMR168, 11 rue Pierre et Marie Curie, 75005, Paris, France}

\date{\today}

\begin{abstract}
    We propose a finite-element method to determine the dynamics of a  fluid interacting with a membrane---a fluid layer equipped with a Helfrich elastic free energy  [W. Helfrich, \href{https://www.degruyterbrill.com/document/doi/10.1515/znc-1973-11-1209/html}{Z.  Naturforschung C, \textbf{28}, 693, (1973)}].
Our analysis is based on a partitioned approach, where the state of the fluid, of the membrane, and of the mesh, are solved for at staggered steps. We show the predicting capabilities of the method by considering a macroscopic example of an elastomer membrane, which bulges as it interacts with air injected through a channel. We demonstrate that our method allows for describing large membrane deformations which are not accessible by perturbative methods.  We discuss the limitations of our method, by characterizing its added-mass instability. Finally, we show how our analysis allows for understanding the physical mechanism of bulge formation, which is due to a combination of elastic stresses and membrane boundary conditions.
\end{abstract}

\maketitle

\section{Introduction}
\label{sec_introduction}
\acresetall

Membranes interacting with a surrounding, bulk fluid are central to a wide range of physical systems. These include biological membranes in contact with the cell aqueous solution or actin network \cite{Bassereau_2018}, cell cortices interacting with the cytoplasm \cite{salbreuxActinCortexMechanics2012}, and soap films in air \cite{Pugh_2016}. For instance,  membrane-fluid interfaces in biology, such as the cell cortex or lipid bilayers, play a key role in cell motility \cite{sensMembraneTensionCytoskeleton2015}, morphogenetic processes \cite{salbreuxActinCortexMechanics2012}, and signaling \cite{diz-munozUseForceMembrane2013}.
Modeling such membrane-fluid interactions presents substantial challenges due not only  to  geometric nonlinearities \cite{worthmullerIRENEFluIdLayeR2025}, but also to the dynamical nature of the fluid spatial domain \cite{doneaChapter14Arbitrary2004}.

Here, we  present a numerical method for the interaction between a fluid and a \acl{mem}---a deformable fluid layer whose material elements can flow, are subject to tension, and whose elastic energy is given by the Helfrich model \cite{helfrichElasticPropertiesLipid1973}. In the latter, the bending energy is proportional to the square of the mean curvature. Originally introduced for lipid bilayers \cite{helfrichElasticPropertiesLipid1973}, this model has become the standard description of the bending elasticity of biological membranes and vesicles \cite{seifertConfigurationsFluidMembranes1997a}, and it describes physical objects across different scales, e.g., thin elastic plates \cite{landauTheoryElasticity1986}.

\subsection{State of the art}\label{sec_state_of_the_art}

A number of existing studies rely on the hypothesis that the \acl{flu} surrounding the \acl{mem} is inertialess.
For instance, Arroyo et al. study the interaction between a \acl{mem} and bulk fluid in the Stokes approximation, for simple membrane geometries that can be described by a single parameter \cite{arroyoRelaxationDynamicsFluid2009i}.
Boedec and coauthors combined  \ac{fe} and \ac{be} methods describe an \acl{mem} interacting with  an inertialess \acl{flu} \cite{boedecIsogeometricFEMBEMSimulations2017}. In there, the membrane surface rheology, such as viscous effects and  flows induced by surface-tension gradients, is neglected.
A \acl{mem} coupled to a \acl{flu} has been recently studied by Zhou and co-authors  with \ac{be} and \ac{fe} methods, under the assumption that the \acl{flu} is inertialess and that \acl{mem}  is  inextensible \cite{zhouModelingSimulationOpen2026}.
Barakat and Shaqfeh  modelled vesicles in a circular tube by means of  \ac{be} methods and lubrication theory, by neglecting \acl{mem} rheology and describing \acl{flu}  as inertialess \cite{barakatStokesFlowVesicles2018}.
Finally, Rodrigues et al., studied the interaction between an inertialess, closed \aclp{mem} and \aclp{flu}  with \ac{fe} methods, by adopting a minimal description for the fluid, which is solely modelled as a volume constraint \cite{rodriguesSemiimplicitFiniteElement2015}.

Other studies go beyond the inertialess approximation, and rely on simplifying assumptions for the \acl{mem}.
For instance, Bonito et al. propose a \ac{fe} model for a \acl{mem} interacting  an \acl{flu},  the latter being in the inertial regime  \cite{bonitoDynamicsBiomembranesEffect2011,bonitoParametricFEMGeometric2010}. In this study the membrane is treated as a purely \acl{ela}, characterized  solely by bending rigidity, area and volume constraints---its fluid features, such as tangential and normal flows, are neglected.
Laadhari extended the analyses above to the interaction   between a \acl{mem} and a non-Newtonian \acl{flu}, in which membrane rheology is still neglected  \cite{laadhariFiniteElementMethodSimulation2023}.
The interaction between a \acl{mem} and a \acl{flu}---both in the inertial regime---has been studied by Barrett and co-authors, \cite{barrettFiniteElementApproximation2016}, under the assumption that the membrane is perfectly inextensible. Such analysis is based on  a fluidic approach, in which the membrane does not obey its own, intrinsic rheology---see below---but it is advected by the surrounding \acl{flu}. As a result, the membrane tangential and normal velocities are not determined as the result of the surfacial \ac{ns} equations, but they match the respective components of the \acl{flu} velocity.

\vspace{0.5cm}\\

In this study, we propose a novel \ac{fe} study of the interaction between a \acl{flu} and a \acl{mem}. Our approach is based on the \ac{ale} method \cite{malvernIntroductionMechanicsContinuous1969,howellp.AppliedSolidMechanics2008,doneaChapter14Arbitrary2004}.
Our analysis  includes a complete description of the physical features of both the membrane and the fluid,  it can describe  inertial regimes of both, and it accounts for a detailed description of membrane elasticity, rheology, and extensibility. Unlike fluidic approaches where the membrane dynamics is determined by the fluid, here both the flow and shape of the membrane are determined by intrinsic physical evolution equations. In addition, our study can account for both closed and open surfaces, by leveraging the flexibility of its \ac{fe} formulation \cite{loggAutomatedSolutionDifferential2012}.

We solve for the system dynamics by  implementing a partitioned solution-scheme \cite{felippaPartitionedAnalysisCoupled2001}, applied to the membrane, the fluid, and the fluid domain. As for the membrane shape, we generalize the \ac{al} gauge---previously proposed to describe the steady state of membrane tubes \cite{derenyiFormationInteractionMembrane2002}---to the system dynamics, by means of a time-dependent gauge fixing. Overall, this implementation allows for handling strongly turbulent membrane and fluid flows, large membrane deformations, and membrane-shape overhangs \cite{worthmullerIRENEFluIdLayeR2025}.\vspace{0.5cm}\\

The paper is structured as follows. In \cref{sec_results} we discuss the \ac{ale} method, and present two illustrative examples: The interaction between a fluid and a rigid body,  \cref{sec_fl_rigid}, and that between  the fluid and the elastic body, \cref{sec_fl_elastic}. In \cref{sec_fl_mem} we present our approach for the interaction between a fluid and a \acl{mem}. In particular, in \cref{sec_fl_mem_mem,sec_fl_mem_dom,sec_fl_mem_flu}, we discuss the splitting scheme and gauge-fixing procedure for the membrane, fluid, and fluid domain, and the temporal discretization is detailed in \cref{sec_temp_discretization_mem}. Finally, \cref{sec_discussion} is devoted to the discussion of the results, and future directions. All used acronyms  are listed in \cref{sec_acronyms}.
\section{Results}\label{sec_results}

Following the general \ac{ale} framework \cite{doneaChapter14Arbitrary2004}, for all  systems that we will consider we  introduce the \ac{ref} and \ac{cur} configuration \cite{kamenskyLectureNotesMAE,landauTheoryElasticity1986,slaughterLinearizedTheoryElasticity2002}, see \cref{fig_ref_curr_rigid}.
All quantities relative to the \ac{ref} and \ac{cur} configuration will be denoted by the superscript \ac{ref} and \ac{cur}, respectively.
In the \ac{ref} and \ac{cur} configuration, the region occupied by the fluid is described by Cartesian coordinates $\bxr$, $\bxc$, respectively.

The \ac{ref} and \ac{cur} configurations are related as follows \cite{kamenskyLectureNotesMAE}, see \cref{fig_ref_curr_rigid}. At  time $t$, the point $\bxc$  corresponds to a point $\bxr$ in the \ac{ref} configuration through the deformation field \cite{landauFluidMechanics1987}
\be\label{eq_def_u}
\budom^t(\bxr) \equiv \bxc - \bxr.
\ee
The deformation-gradient tensor is defined as \cite{landauTheoryElasticity1986}
\be
\label{eq_def_F}
F_{\alpha \beta}(\bu) \equiv \delta_{\alpha \beta} + \pderr{u^ \alpha}{\beta}.
\ee
In what follows, Greek indices $\alpha, \beta, \cdots$ will be used to denote vectors and tensors in  two-dimensional Euclidean space---their position as upper or lower indices is thus immaterial \cite{marchiafavaAppuntiDiGeometria2005,landauTheoryElasticity1986}.
We will characterize  the mapping between $\bxc$ and $\bxr$ configurations in \cref{eq_def_u} through the fields
\begin{align}\label{eq_def_phi}
        \bxc & = \bm{\varphi}^t(\bxr), \\
        \label{eq_def_psi}
        \bxr & = \bm{\psi}^t(\bxc).
\end{align}

We denote the \ac{cur} fluid velocity and surface tension, i.e., the negative of the pressure, by $\vfc(\bxc)$ and $\sigmafc{}(\bxc)$, respectively.
For each field $\bvfc{}({\bxc}), \sigmafc{}(\bxc)$, we introduce its counterpart in the \ac{ref} configuration:
\begin{align}
        \label{eq_vf_ref}
        \bvfr{}(\bxr, t)    & \equiv \bvfc{}(\bm{\varphi}^t(\bxr), t),      \\
        \label{eq_sigmaf_ref}
        \sigmafr{}(\bxr, t) & \equiv  \sigmafc{}( \bm{\varphi}^t(\bxr), t).
\end{align}

The solution of the interaction dynamics between the fluid and the rigid body, elastic body or membrane, is  involved because the fluid domain has a dynamical, moving boundary, cf. \cref{fig_ref_curr_rigid,fig_ref_curr_ela,fig_ref_cur_mem}, respectively. Following the \ac{ale} procedure, we will  bridge between the Lagrangian   description used to describe the rigid body, elastic body or membrane, and the Eulerian description of the fluid  \cite{kamenskyLectureNotesMAE,doneaChapter14Arbitrary2004}. To achieve this, we will consider the equations of motion for the rigid body, elastic body or membrane and the fluid---which are naturally formulated the \ac{cur} configuration---and  rewrite them in the \ac{ref} configuration \cite{kamenskyLectureNotesMAE}.

We will first discuss this procedure in the illustrative examples of interaction between a \acl{flu} and a \acl{bod} and that between a \acl{flu} and an \acl{ela}, and then consider  a fluid  interacting with a \acl{mem}.

\subsection{\Acl{flu} and \acl{bod}}\label{sec_fl_rigid}

In this Section, we will discuss the interaction between a  fluid and the simplest structure---a  the rigid body which is only allowed to turn about one point. Only the main results will be presented here; details can be found in \cref{sec_rigid_body_body,sec_rigid_body_fl,sec_eq_fluid_domain}, respectively, and the temporal discretization is discussed in \cref{sec_temp_discretization_rig}.

As shown in  \cref{fig_ref_curr_rigid}, in the \ac{ref} configuration the  axis of the ellipse, i.e., the  body, lies parallel to the $\xc[1]$ axis, while in the \ac{cur} configuration, such axis forms an angle  $\theta$ (red arc) with respect to such axis.

\begin{figure*}[t]
        \centering
        \includegraphics[width=\textwidth]{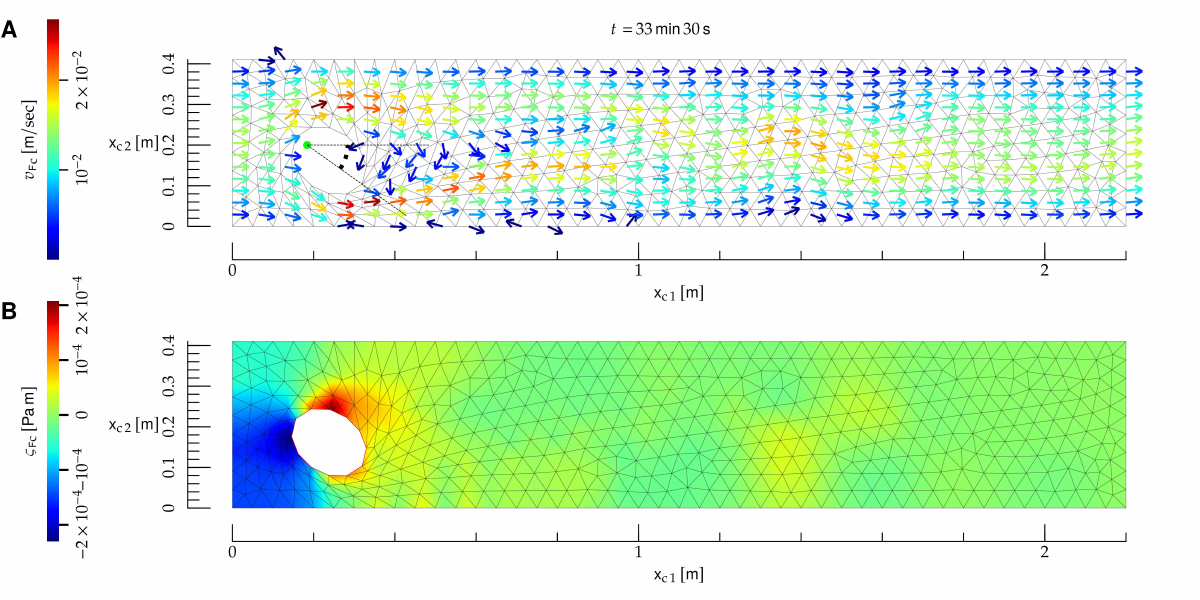}
        \caption{
                \label{fig_fluid_rigid}
                Interaction between a  \acf{flu} and a  \acl{bod}. The \acl{flu} material properties roughly correspond to air, and those of the body to an aerogel.  \plab{A} Temporal snapshot of the mesh (black lines) and \acl{flu} velocity field (arrows) at the \acl{cur} time $t$, shown on top. The direction of the velocity field is indicated by the arrows, and its norm by their color; the direction of arrows with velocity close to zero is not defined. The body (ellipse) is allowed to pivot about its left focal point (green dot); the related angle with respect to the $\xc[1]$ direction is denoted by a black dashed  arc.  The Reynolds number is $R \sim 2 \times 10^2$.
                \plab{B} Color map of the surface tension at the \acl{cur} time $t$.
        }
\end{figure*}

An example of the system dynamics is shown in \cref{fig_fluid_rigid}, where we study the interaction between air (the fluid) and an aerogel (the rigid body)---an ultra light solid material involving a large porous structure \cite{kistlerCoherentExpandedAerogels1931}. The channel  dimension are $L = \parameter[2.2] {\met}$, $h =  \parameter[0.41]{\met}$, which have been taken from the \ac{feat2d} DFG 2D-3 benchmark example for a fluid flow around a cylinder \cite{blumFEAT2DFiniteElement1995,schaferBenchmarkComputationsLaminar1996}, see  \cref{fig_ref_curr_rigid}. Given that the materials, such as the fluid and the  body, are two dimensional, in what follows we will estimate their parameter values by considering their three-dimensional counterparts, and assume a unit thickness in the out-of-plane direction.
The fluid parameters  have been estimated from the physical properties of air: $\rhof = 1 \, \kg / \met^2$, $\etaf = \parameter[][-5]{\pas, \met, \second}$ \cite{hilsenrath1955tables}. The  inflow velocity profile is chosen as a Poiseuille flow
\be\label{eq_inflow_profile}
\vfin(\bxc) = \frac{6 v_0}{h^2} \xc[2] (\xc[2] - h),
\ee
with $v_0 = 1 \, \units{\met}[\second]$. The  mesh of the fluid domain deforms in time, and its rigidity  is determined by the stiffness exponent $\zeta$ \cite{kamenskyLectureNotesMAE}, which  has been chosen to be $\zeta = 3$, see \cref{sec_fl_mem_dom} for details.  Finally,  the rigid body is an ellipse with semi-axes $a = \parameter[0.1]{\met}$, $b = \parameter[0.075]{\met}$,  center located at  $\bm{c} = (0.25 \, \met, 0.2\, \met)$, and moment of inertia $I = \parameter[][-3]{\kg, \met^2}$, which corresponds to an aerogel density of $\rhob = \parameter[6]{\kg}[\met^3]$ \cite{niculescuUpdatedOverviewSilica2024}. In \cref{sec_fluid_rigid_timescale} we discuss a consistency check of the \ac{ale} scheme above, showing that the oscillation period of the body matches the prediction obtained with scaling arguments---see \cref{sec_fluid_rigid_timescale}.

\subsection{\Acl{flu} and \acl{ela}}\label{sec_fl_elastic}

In this Section, we will build on the analysis of \cref{sec_fl_rigid} and put the fluid into interaction with a more complex physical object---an \acl{ela}; only the main results will be presented here.

The \ac{ref} and \ac{cur} configurations are shown in \cref{fig_ref_curr_ela}. We allow the  elastic-body boundary $\pomcircouteq$ to deform, but the inner body boundary $\pomcircineq$ is pinned to the $\xc[1]$, $\xc[2]$ plane.
The equations of motion are reported in \cref{app_sec_fl_elastic_equations_of_motion}, and the system dynamics is discussed in \cref{app_sec_fl_elastic_dynamics}.

\begin{figure*}
    \centering
    \includegraphics[width=\textwidth]{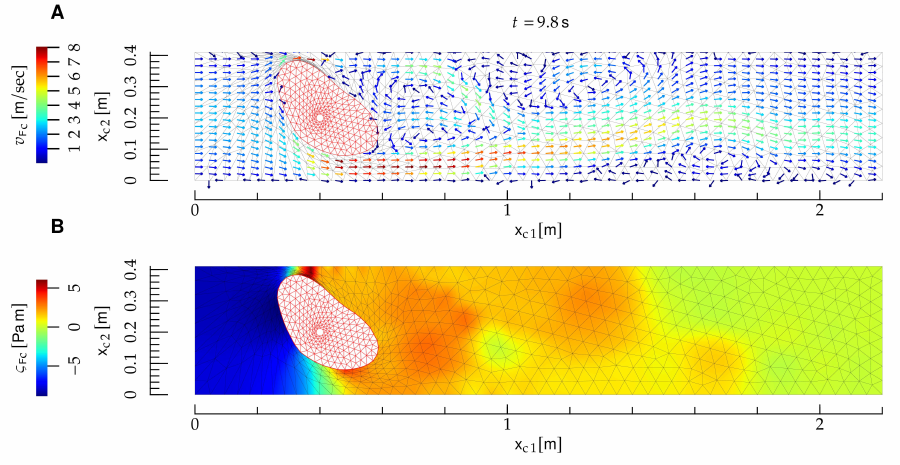}
    \caption{
        \label{fig_fluid_ela}
        Interaction between a \acf{flu} and an \acl{ela} (a hydrogel). The notation is the same as in \cref{fig_fluid_rigid}, and the body is depicted as a red mesh.  The elastic-body deformation is not symmetric under the mirror symmetry $\xc[2] \rightarrow h-\xc[2]$, where $h$ is the height of the \acl{flu} domain, because the center of the body is not located at $\xc[2] = h/2$. The Reynolds number is $R \sim 2 \times 10^2$.
    }
\end{figure*}

An example of such  dynamics  is depicted in \cref{fig_fluid_ela}, where we show the interaction between a fluid and an hydrogel (the elastic body).
Here, $\pomcircouteqr$ is an ellipse with semi-axes $a = 0.2\, \met$, $b = 0.1\, \met$ and center located at  $\bm{c} = (0.4 \, \met, 0.2\, \met)$, and $\pomcircineqr$ has radius $r = \parameter[1.25][-2]{\met}$ and center located at $\bm{c}$.
Channel dimensions are the same as in \cref{fig_fluid_rigid}, and fluid parameters have been taken from the \ac{feat2d} DFG 2D-3 benchmark example for a fluid flow around a cylinder \cite{blumFEAT2DFiniteElement1995,schaferBenchmarkComputationsLaminar1996}: $\rhof = 1 \, \kg/\met^2$, $\etaf = 10^{-3}\, \pas \, \met \, \second$.
The body density is $\rhoela = 10^3 \,\kg/\met^2$   \cite{xuConcurrentStiffeningSoftening2023}. The storage modulus, which approximates the shear modulus $\mu$ \cite{REHAGE1974161} is $ \sim  10 \, \kg/\second^2$ \cite{lichtSynthetic3DPEGAnisogel2019}.
Given an elastic-body Poisson ratio \cite{landauTheoryElasticity1986} of $\poissonratio \sim 0.49$ \cite{chippadaSimultaneousDeterminationYoungs2010}, we obtain $K \sim 5 \times 10^2 \, \kg/\second^2$ from the relation $K = 2 \mu(1+\poissonratio)/[3(2\poissonratio-1)]$ \cite{landauTheoryElasticity1986}.
The fluid inflow velocity profile is  a Poiseuille flow, \cref{eq_inflow_profile}, with $ v_0 = 1\, \met / \second$. Finally, the fluid-domain mesh stiffness exponent has been chosen to be $\zeta = 3$ \cite{kamenskyLectureNotesMAE}.

\subsection{\Acl{flu} and \acl{mem}}\label{sec_fl_mem}

In this Section, we will build on the analysis of \cref{sec_fl_elastic}, and complexify the structure of the body with which the fluid interacts. We consider, instead of an elastic material, a Helfrich membrane. The latter is a deformable fluid layer whose elements can flow, equipped with an elastic, bending energy given by the Helfrich model \cite{helfrichElasticPropertiesLipid1973,derenyiFormationInteractionMembrane2002}.

We will consider the  geometry depicted in \cref{fig_ref_cur_mem}. A vertical channel with mirror symmetry about its mid axis is considered. The left boundary of the channel is \pomin{}, and the right boundary \pomout{} lies on the symmetry axis. The fluid is injected  on \pombottom, from bottom to top. The membrane lies on the top boundary of the fluid, \pomtop{}, which  coincides with the membrane manifold \pomline{}. Finally, we denote by $\om$ the region enclosed by the boundaries, and all the quantities above will be considered in both the \ac{ref} and \ac{cur} configuration.

The geometry of \cref{fig_ref_cur_mem} may represent the fluid being confined in a  reservoir, lying below the channel, where the channel is a vertical neck of the reservoir, covered on top by the membrane.

\subsubsection{Membrane motion}
\label{sec_fl_mem_mem}

\begin{figure}
    \centering
    \includegraphics[width=\columnwidth,height=0.55\textheight,keepaspectratio]{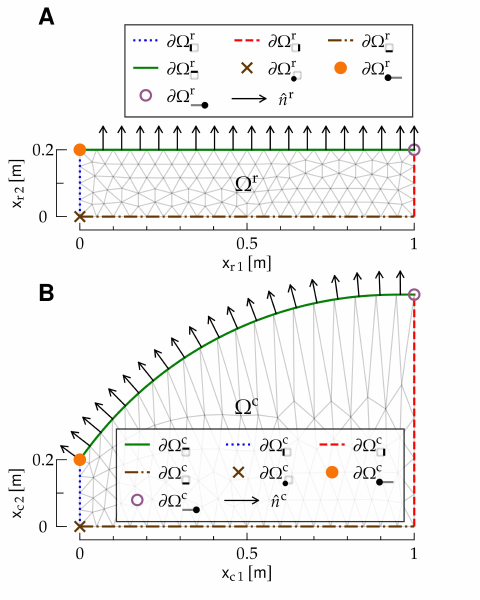}
    \caption{
        \label{fig_ref_cur_mem}
        \Acf{ref} and \acf{cur} configuration for a fluid and a \acl{mem}.
        The Figure  shows a vertical channel with mirror symmetry about its mid axis (red dashed line), see \cref{fig_ref_curr_rigid}. The right, mirrored part of the channel is not shown, cf. \cref{fig_fluid_mem_mem}A.
        \plab{A} \ac{ref} configuration.
        The top edge of the \acl{flu} domain, $\pomtopeqr$, coincides with the one-dimensional  manifold $\omlineeqr$ of the membrane.
        We also show         the other boundaries of the fluid domain, $\pomineqr$, $\pomouteqr$, $\pombottomeqr$, the bottom-left point of the fluid domain $\pomleftbottomeqr$, , and the membrane boundaries, $\pomlineineqr$ and $\pomlineouteqr$.
        The \ac{ref} unit normal $\normalr{}$ is also shown; its  plotted length has been adjusted for clarity.
        \plab{B} Same as \textbf{A}, for the \ac{cur} configuration.
        The membrane profile is purposely, strongly deformed without remeshing, so as to illustrate mesh deformation.
    }
\end{figure}

Both the physical nature and mathematical description of a Helfrich membrane are significantly more complex than that of an elastic body. First, unlike an elastic body, the membrane has a fluid structure: Its material elements can  flow tangentially to the membrane shape, which also moves subjected to  bending-rigidity forces, and to the force exerted on it by the fluid.
While the tangential motion is naturally suited to an Eulerian description \cite{landauFluidMechanics1987}, the shape motion---reminiscent of that of an elastic body \cite{helfrichElasticPropertiesLipid1973}---is best described with a Lagrangian approach \cite{lacarbonaraNonlinearStructuralMechanics2013}.
Also, such shape motion needs to be connected to the Lagrangian approach used to describe the fluid domain. Second, the membrane is described by fourth-order \acp{pde} \cite{zhong-canBendingEnergyVesicle1989}. From  the mathematical and numerical point of view, such \acp{pde} are significantly more complex than the second-order \acp{pde} which describe an elastic body \cite{worthmullerIRENEFluIdLayeR2025}. On top of this, from the physical standpoint, such fourth-order \acp{pde} require more  \acp{bc} than the second-order ones for an elastic body: Such \acp{bc} may specify a number of physical features of the membrane system, such as forces, positions, torques, or stresses.
Third, the geometrical description of the membrane requires fixing an appropriate \textit{gauge}.
In fact, a general parametrization of the membrane shape is given by two functions---the two Cartesian components of the membrane shape $\bxmemc(\xcoord^1)$. Both of these functions  depend on the curvilinear coordinate $\xcoord^1$ which is defined in the \ac{ref} domain of the membrane, $\pomlineeqr$, see \cref{fig_ref_cur_mem}---note that here $\xcoord^i$ differs from the Cartesian coordinate $\xcartesian^\alpha$.
Despite this pair of functions, a single \ac{pde}  determines the membrane shape \cite{zhong-canBendingEnergyVesicle1989}.
To understand this redundancy, consider the  stretching field $\nu>0$, defined by

\begin{align}\label{eq_intro_nu}
    \nu ^ 2   & \equiv \bevec{1}   \cdot \bevec{1}, \\
    \label{eq_def_e}
    \bevec{i} & \equiv \pder{{\bxmemc}}{\xcoord^i}.
\end{align}
Two  parametrizations with different $\nu$s can yield the same physical shape. Such  redundancy  may be removed by a gauge fixing, i.e., by adding a constraint on $\bxmemc$, or its derivatives.
Two common gauge choices are the Monge  \cite{hsiungFirstCourseDifferential1981,desernoFluidLipidMembranes2015}, or the arc-length parametrization \cite{derenyiFormationInteractionMembrane2002}, in which one sets
\be
\label{eq_arc_length_gauge}
\nu = 1.
\ee
The arc-length gauge has been used to describe, for instance, the steady state for membrane tubules \cite{derenyiFormationInteractionMembrane2002}. However, while a static constraint such as \crefs{eq_arc_length_gauge} is suited to  static shapes, it cannot describe the  membrane dynamics, such as the one of a membrane interacting with a fluid. In fact, as the membrane is deformed, it will also stretch, and \cref{eq_arc_length_gauge} must be replaced by a time-dependent  gauge---see below.

We describe the membrane with the Lagrangian approach, along the lines of the description of the fluid domain in \cref{sec_fl_rigid} and an elastic body  in \cref{sec_fl_elastic}.  The membrane displacement field is
\be\label{eq_u_mem}
\bumem(\xcoord^1) \equiv \bxmemc(\xcoord^1) - \bxmemr(\xcoord^1).
\ee
At the beginning of the dynamics, we choose the \ac{ref} configuration as a straight line
\be
\label{eq_def_x_ref}
\bxmemr(\xcoord^1) \equiv (\xcoord^1, h), \, 0 \leq  \xcoord^1 \leq L,
\ee
see \cref{fig_ref_cur_mem}A.
The mesh will then be deformed with time: As the deformation becomes large, we will remesh the fluid domain. The \ac{cur} configuration right before remeshing will then become the \ac{ref} configuration after remeshing, which will be, in general,  different from \crefs{eq_def_x_ref}.

In addition, we introduce the  angle $\psi$ formed by the tangent to the membrane with the $\xcartesian^1$ axis \cite{derenyiFormationInteractionMembrane2002}, defined by
\begin{align}
    \label{def_nu_psi_1}
    \evec{1}{1} & = \nu \cos \psi, \\
    \label{def_nu_psi_2}
    \evec{1}{2} & =-\nu \sin \psi.
\end{align}
Finally, the membrane metric tensor is defined by  \cite{marchiafavaAppuntiDiGeometria2005}
\be\label{eq_def_g}
g_{ij} \equiv \bevec{i} \cdot \bevec{j}.
\ee
In what follows, Latin indices $i, j, k, \cdots$ will be used to denote vectors and tensors related to the membrane manifold; their position as upper and lower indexes is thus important \cite{marchiafavaAppuntiDiGeometria2005}. Because the differential manifold of the membrane is one dimensional, indexes $i, j, k, \cdots$ may take only one value, i.e., $i=j=k= \cdots=1$. However, in formal relations such as \cref{eq_def_g}, we will use the general notation $i, j, k, \cdots$  for the sake of  generality, and for potential extensions of such relations to two-dimensional manifolds \cite{desernoFluidLipidMembranes2015}.
\vspace{0.5cm}\\

In order to work out the  dynamical equations for the membrane, let us derive the force exerted by the fluid on the membrane. The force exerted on a line element $\dint{l_\current}$ of the membrane  is \cite{landauFluidMechanics1987}
\be
\label{def_f_fluid_mem}
f_{\fluid\, \alpha}(\bsigmastressc, \nu, \psi) = - \sigmastressc{\alpha}{\beta} \normalc{\beta}(\nu, \psi),
\ee
where
\be
\label{eq_strees_tensor_fluid}
\sigmastressc{\alpha}{\beta} \equiv \sigmafc{} \delta_{\alpha \beta}+ \eta_\fluid\left( \pderc{ \vfc^\alpha}{\beta} +\pderc{ \vfc^\beta}{\alpha}\right)
\ee
is the fluid stress tensor \cite{landauFluidMechanics1987} and $\normalc{}$ the unit normal to $\pomtopeqc$ pointing outside $\omc$, see \cref{fig_ref_cur_mem}B.  In \cref{def_f_fluid_mem}, $\bsigmastressc$ is evaluated at the \ac{cur} coordinate $\bxmemc \in \pomtopeqc$ corresponding to $\xcoord^1$.
The components of the force \crefs{def_f_fluid_mem} tangential and normal to the membrane are \cite{desernoFluidLipidMembranes2015}
\begin{align}
    \label{def_force_tan}
    f_{\fluid\, \text{t}}^i(\bsigmastressc, \nu, \psi) & = g^{ij}(\nu) \evec{j}{\alpha}(\nu,\psi) f_\alpha(\bsigmastressc, \nu, \psi), \\
    \label{def_force_norm}
    \fflun (\bsigmastressc, \nu, \psi)                 & = \normalc{\alpha}(\nu, \psi) f_\alpha(\bsigmastressc, \nu, \psi),
\end{align}
where in \cref{def_force_tan} we used the feature that the metric tensor depends on $\nu$ only, see \cref{eq_def_g,eq_intro_nu}.

The equations of motion for the membrane  are \cite{worthmullerIRENEFluIdLayeR2025,al-izziShearDrivenInstabilitiesMembrane2020,arroyoRelaxationDynamicsFluid2009i}

\bw
\begin{align}
    \label{eq-dyn-continuity}
    \nab_i \vmem{i} - 2 H \wmem                                                                              =               & 0,                                                                                                                                                                      \\
    \label{eq-dyn-v}
    \rhom \left( \pder{\vmem{i}}{t}+ \vmem{j} \nab_j \vmem{i}  - 2 \vmem{j} \wmem\, b^i_j - \wmem \nab^i \wmem \right)     = & \fmemt^i + f_{\fluid\, \text{t}}^i,                                                                                                                                     \\
    \label{eq-dyn-w}
    \rhom \left[ \pder{\wmem}{t} + \vmem{i} ( \vmem{j} b_{ji} + \nabla_i \wmem  ) \right]                               =    & f_\lapl   +  f_\kap +                                                         \fetamemn+ \fflun                                                                       , \\
    \label{eq-dyn-dX}
    \pder {\bumem}{t}                                                                                            =           & \wmem \, \normalc{}{},                                                                                                                                                  \\
    \label{eq-dyn-dX1}
    \pder {\umem^1}{x^1}                                                                                            =        & - \pder{\xmemr{1}}{\xcoord^1} + \nu \cos \psi,                                                                                                                          \\
    \label{eq-dyn-dX2}
    \pder {\umem^2}{x^1}                                                                                               =     & - \pder{\xmemr{2}}{\xcoord^1} -\nu \sin \psi,
\end{align}
\ew
and they hold for $\xcoord^1 \in \omlineeqr$.
Here, $\vmem{i}$ and $w$ are the membrane tangential and normal velocity, respectively, $\nab$ is the covariant derivative related to the Levi-Civita connection associated to $g$, and $H$ the mean  curvature.
We refer to \cite{desernoFluidLipidMembranes2015,worthmullerIRENEFluIdLayeR2025,marchiafavaAppuntiDiGeometria2005,arroyoRelaxationDynamicsFluid2009i} for details on the differential-geometry formulation.
We will now discuss each line in \cref{\eqsmembrane}, and  set

\begin{align}
    \label{eq_def_f_lap}
    f_\lapl (\sigmam{}, \nu,\psi)               & \equiv  2\, \sigmam{} H ,                                   \\
    \label{eq_def_f_kap}
    f_\kap (\nu,\psi)                           & \equiv  -2\kap [ \nab_i \nab^i H + 2 H (H^2-K)],            \\
    \label{eq_def_fmemt}
    \fmemt^i(\vmem{}, \wmem,\sigmam{},\nu,\psi) & \equiv - \nab_j \Pi^{ij}(\vmem{}, \wmem,\sigmam{},\nu,\psi) \\
    \label{eq_def_f_eta_n}
    \fetamemn      (\vmem{}, \wmem, \nu, \psi)  & \equiv
    2 \etam [ (\nab^i \vmem{j})b_{ij} - 2 \wmem (2 H^2 - K)                                                                 ]  ,
\end{align}
where
\begin{align}
    \label{eq_def_Pi}
    \Pi^{ij}(\vmem{}, \wmem,\sigmam{},\nu,\psi) & \equiv    -   g^{ij}\sigmam{} - 2 \etam d^{ij}(\vmem{}, \wmem,\nu,\psi) , \\
    \label{eq_def_d}
    d^{ij}(\vmem{}, \wmem,\nu,\psi)             & \equiv  \frac{1}{2}( \nab^i \vmem{j}  + \nab^j \vmem{i}) - \wmem b^{ij}.
\end{align}
In \cref{eq_def_Pi} $\Pi$, is the momentum-flux density tensor \cite{landauFluidMechanics1987}, which is given by a line-tension term and by a viscous term, where the latter is expressed in terms of the rate-of-deformation tensor $d$ \cite{al-izziShearDrivenInstabilitiesMembrane2020}. Finally, in \cref{eq_def_d}, $b$ is  the second fundamental form \cite{marchiafavaAppuntiDiGeometria2005,desernoNotesDifferentialGeometry2004}.

In \cref{\eqsmembrane}, $\rhom$, $\sigmam{}$, $\etam$ and $\kap$ are, respectively, the density, line tension, viscosity, and bending rigidity of the membrane.
\Cref{eq-dyn-continuity} enforces mass conservation \cite{landauFluidMechanics1987}.
\Cref{eq-dyn-v,eq-dyn-w} are the tangential and normal components of the \ac{ns} equations, respectively.
The term $\fmemt$ in \cref{eq-dyn-v} represents the force on membrane elements induced by gradients of line tension and viscous effects, cf. \cref{eq_def_fmemt}, while $f_{\fluid\, \text{t}}$ is the force exerted on membrane elements by the fluid.
In \cref{eq-dyn-w}, $f_\lapl$ constitutes Laplace pressure \cite{degennesCapillarityWettingPhenomena2004,landauFluidMechanics1987},  $f_\kap$  the resistance of the membrane to bending \cite{helfrichElasticPropertiesLipid1973}, cf. \cref{eq_def_f_kap},  while $\fetamemn$ and $\fflun$ are  the  normal components of the viscous force resulting from in-membrane flows \cite{arroyoRelaxationDynamicsFluid2009i}, and of the force exerted by the fluid on the membrane, respectively.
\Cref{eq-dyn-dX} is the time-evolution equation for the membrane shape, which moves according to the normal $\normalc{}$, and it has been obtained from  \cref{eq_u_mem}.
We observe that the membrane shape velocity, $\partial \bumem / \partial t$, coincides with the normal displacement velocity of the membrane, but  it is independent of the tangential flow velocity $\bvmem{}$ of membrane elements along the membrane shape.
This implies that the membrane shape and fluid motions allow for a slip \cite{landauFluidMechanics1987}, while  being geometrically conforming \cite{kamenskyLectureNotesMAE}.
Finally, \cref{eq-dyn-dX1,eq-dyn-dX2} re-express the definitions of $\nu$ and $\psi$, and  they result from \cref{def_nu_psi_1,def_nu_psi_2,eq_def_e,eq_u_mem}, and from the fact that the \ac{ref} shape $\bxmemr$ is independent of time.

For future reference, we define the inertial terms \cite{landauFluidMechanics1987} in  the \ac{lhs} of \cref{eq-dyn-w} as follows:
\begin{align}
    \label{eq_def_I_v}
    I_v & \equiv  \rhom \vmem{i}  \vmem{j} b_{ji}, \\
    \label{eq_def_I_w}
    I_w & \equiv  \rhom \vmem{i} \nabla_i \wmem.
\end{align}

We choose the following \acp{bc} for \cref{\eqsmembrane}:
\begin{align}
    \label{bc_fl_mem_mem_1}
    \normalmem_j \Pi^{ij}     & = 0 \text{ on } \pomlineineqr,  \\
    \label{bc_fl_mem_mem_2}
    \vmem{i}                  & = 0 \text{ on } \pomlineouteqr, \\
    \label{bc_fl_mem_mem_3}
    \wmem                     & = 0 \text{ on } \pomlineineqr,  \\
    \label{bc_fl_mem_mem_4}
    \nab_i \wmem              & = 0 \text{ on } \pomlineouteqr, \\
    \label{bc_fl_mem_mem_5}
    \sigmam{}                 & = 0 \text{ on } \pomlineineqr,  \\
    \label{bc_fl_mem_mem_6}
    \bumem                    & = 0 \text{ on } \pomlineineqr,  \\
    \label{bc_fl_mem_mem_7}
    \umem^1                   & = 0 \text{ on } \pomlineouteqr, \\
    \label{bc_fl_mem_mem_8}
    \pder{\umem^2}{\xcoord^1} & = 0 \text{ on } \pomlineouteqr.
\end{align}

\Cref{bc_fl_mem_mem_3,bc_fl_mem_mem_6} enforce the condition that the left extremity of the membrane shape is clamped on the channel wall \pomin.
\Cref{bc_fl_mem_mem_1,bc_fl_mem_mem_5} imply that the \ac{le} extremity of the membrane is not subjected to  external forces \cite{landauFluidMechanics1987}, i.e., it can freely flow tangentially.
\Cref{bc_fl_mem_mem_4,bc_fl_mem_mem_7,bc_fl_mem_mem_8,bc_fl_mem_mem_2} enforce the mirror symmetry about \pomout.
Namely, \cref{bc_fl_mem_mem_2} ensures that  $\bvmem{}$  vanishes on \pomlineout, because a nonzero value  would break the axial symmetry. The condition \crefs{bc_fl_mem_mem_7} follows the same lines. In addition, the  symmetry requires that scalar fields and vertical components of vector fields in $\om$ have zero slope with respect to $\xcoord^1$ at \pomlineout{}, see \cref{bc_fl_mem_mem_4,bc_fl_mem_mem_8}.

\subsubsection{Fluid-domain motion}
\label{sec_fl_mem_dom}

We will determine $\budom$ as the solution of a \ac{bvp} given by the following elastic model \cite{kamenskyLectureNotesMAE,marsdenMathematicalFoundationsElasticity1994}
\be\label{eq_mesh_fluid_rigid}
\pder{\Sdom_{\alpha \beta}(\budom)}{\xr[\beta]} = 0 \text{ in } \omr,
\ee
where the elastic stress tensor $\Sdom$ is given by
\be
\label{eq_def_S}
\Sdom_{\alpha \beta} \equiv F_{\alpha \gamma} T_{\gamma \beta},
\ee
$F$ is defined in \cref{eq_def_F}, and
\begin{align}
    \label{eq_ela_1}
    T_{\alpha \beta}  \equiv & \kel(\budom) E_{\gamma \gamma} \delta_{\alpha \beta} +\newlinenn
                             & 2 \muel(\budom) \left(E_{\alpha \beta} - \frac{1}{2} \delta_{\alpha \beta} E_{\gamma \gamma}\right), \\
    \label{eq_ela_2}
    E_{\alpha \beta}  \equiv & \frac{1}{2}\left( F_{\gamma \alpha}F_{\gamma \beta} - \delta_{\alpha \beta}\right),                  \\
    \label{eq_ela_3}
    \kel(\bu)         \equiv & \frac{1}{[\det(F(\bu))]^\zeta},                                                                      \\
    \label{eq_ela_4}
    \muel(\bu)        \equiv & \frac{1}{[\det(F(\bu))]^\zeta}.
\end{align}
In \cref{eq_ela_3,eq_ela_4}, the bulk and shear modulus, $\kel$ and $\muel$, respectively, depend on the deformation-gradient tensor \crefs{eq_def_F}. In the absence of this dependence, the elastic model above would be unstable under local mesh compression \cite{kamenskyLectureNotesMAE}, such as the one shown below the rigid body in \cref{fig_ref_curr_rigid}B. In fact, the model Lagrangian would not penalize configurations with  small $\det F$. The inclusion of the dependence \crefs{eq_ela_3,eq_ela_4} on $\det F$  allows for penalizing these configurations---the larger the exponent $\zeta > 0$, the stronger the penalty.

The equation of motion for $\budomdot$ is obtained by deriving \cref{eq_mesh_fluid_rigid} with respect to time, and using the fact that \ac{ref} coordinates are time independent:
\be
\label{eq_mesh_fluid_rigid_dot}
\pder{}{\xr[\beta]}\der{\Sdom_{\alpha \beta}(\budom)}{t}  = 0 \text{ in } \omr.
\ee
We impose the following \acp{bc} for  \cref{\eqsdom}:
\begin{align}
    \label{bc_fl_mem_dom_1}
    \budom                    & = \bzero \text{ on }\pomineqr \cup \pombottomeqr, \\
    \label{bc_fl_mem_dom_2}
    \udom^1                   & = 0 \text{ on } \pomouteqr,                       \\
    \label{bc_fl_mem_dom_3}
    \pder{\udom^2}{\xr[1]}    & = 0 \text{ on }\pomouteqr,                        \\
    \label{bc_fl_mem_dom_4}
    \budom                    & = \bumem \text{ on }\pomtopeqr,                   \\
    \label{bc_fl_mem_dom_5}
    \budomdot                 & = \bzero \text{ on }\pomineqr \cup \pombottomeqr, \\
    \label{bc_fl_mem_dom_6}
    \udomdot^1                & = 0 \text{ on } \pomouteqr,                       \\
    \label{bc_fl_mem_dom_7}
    \pder{\udomdot^2}{\xr[1]} & = 0 \text{ on }\pomouteqr,                        \\
    \label{bc_fl_mem_dom_8}
    \budomdot                 & = \bumemdot \text{ on }\pomtopeqr.
\end{align}
\Cref{bc_fl_mem_dom_1} ensures that the fluid domain is not deformed on $\pomineqr \cup \pombottomeqr$, cf. \cref{bc_ela_1}. \Cref{bc_fl_mem_dom_2,bc_fl_mem_dom_3} follow from axial symmetry.
\Cref{bc_fl_mem_dom_4} derives from the requirement that  fluid-domain and membrane deformations are conforming \cite{kamenskyLectureNotesMAE}: In \cref{bc_fl_mem_dom_4}, $\budom$ is evaluated at a reference coordinate $\bxr \in \pomtopeqr$, and $\bumem$ is evaluated at the curvilinear coordinate $\xcoord^1$ such that $\bxmemr(\xcoord^1) = \bxr$. Finally, the  \acp{bc} for $\budomdot$, \cref{bc_fl_mem_dom_5,bc_fl_mem_dom_6,bc_fl_mem_dom_7,bc_fl_mem_dom_8}, are derived along the same lines as \cref{bc_fl_mem_dom_1,bc_fl_mem_dom_2,bc_fl_mem_dom_3,bc_fl_mem_dom_4}.

\begin{figure*}
    \centering
    \includegraphics[width=0.95\textwidth]{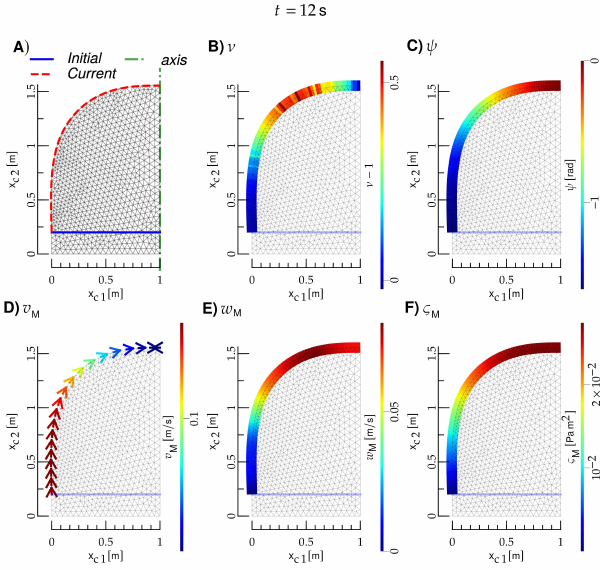}
    \caption{
        Interaction between a fluid  and a \acf{mem}. Here we show the membrane fields.  The fluid is air and the membrane an elastomer.
        \plab{A}  Initial and \acf{cur} configuration of the membrane (solid blue and dashed red curve, respectively). Only the \acl{le} half of the system is shown, the \acl{ri} half is obtained by mirror symmetry from the \acl{le} half through the symmetry axis (green dotted and dashed line).
        \plab{B} and \plab{C}: Membrane stretch field and tangent angle, respectively.
        \plab{D}, \plab{E} and \plab{F}: Membrane tangential velocity, normal velocity and tension, respectively.
        All panels refer to the same instant of time, shown on top, and they also display the mesh. In panels \textbf{B}-\textbf{F}, the initial  membrane configuration is also shown (dim blue line).
        \label{fig_fluid_mem_mem}
    }
\end{figure*}

\begin{figure*}
    \centering
    \includegraphics[width=0.7\textwidth]{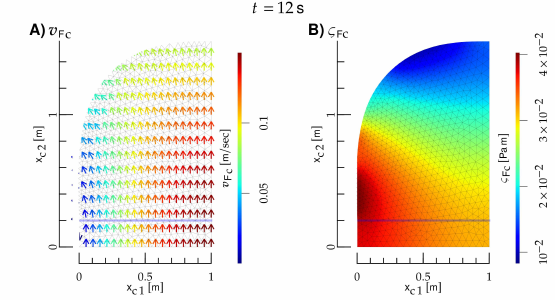}
    \caption{
        Interaction between a \acf{flu}  and a \acl{mem}. Here we show the fluid fields. The solution and notation is the same as in \cref{fig_fluid_mem_mem}.
        \plab{A} and \plab{B}: Fluid velocity and surface tension, respectively. The mesh and  initial membrane configuration (blue line) are also shown.
        \label{fig_fluid_mem_bflu}
    }
\end{figure*}

\subsubsection{\Acl{flu} motion}
\label{sec_fl_mem_flu}

We will now discuss the fluid equations of motion, and present them first in \ac{cur} configuration coordinates, and then transform them in \ac{ref} coordinates.

\paragraph{Equations of motion in  \acl{cur} configuration coordinates}
\label{sec_fl_mem_flu_cur}

In \ac{cur} coordinates, the fluid equations of motion are the \ac{ns} and continuity equations \cite{landauFluidMechanics1987}:

\begin{align}
    \label{eq_fl_rigid_fl_1_current}
    \rho_\fluid \left( \pder{\vfc^\alpha}{t} + \vfc^\beta  \pderc{\vfc^\alpha}{\beta} \right) & =  \pderc{  \sigmastressc{\alpha}{\beta}}{\beta} ,                                               \\
    \label{eq_fl_rigid_fl_2_current}
    \pderc{ \vfc^\alpha }{\alpha}                                                             & =                                                              0                               ,
\end{align}
and they hold for $\bxc \in \omc$.
\Cref{\eqsbfluc} describe the fluid motion in a flat, two-dimensional geometry \cite{landauFluidMechanics1987}---an example of the corresponding  \ac{ns} equations on a curved geometry is given by \cref{eq-dyn-v}.

We enforce the following \acp{bc} for \cref{\eqsbfluc}:
\begin{align}
    \label{bc_fl_membrane_1_current}
    \bvfc{}           & =  \vfbottom  \text{ on } \pombottomeqc,            \\
    \label{bc_fl_membrane_2_current}
    \bvfc{}           & =  \bzero  \text{ on } \pomineqc,                   \\
    \label{bc_fl_membrane_3_current}
    \vfc^1            & =  0 \; \text{ on } \pomouteqc ,                    \\
    \label{bc_fl_membrane_4_current}
    \pderc{\vfc^2}{1} & =  0 \; \text{ on } \pomouteqc ,                    \\
    \label{bc_fl_membrane_5_current}
    \bvfc{}           & = \budomdot  \text{ on } \pomtopeqc,                \\
    \label{bc_fl_membrane_6_current}
    \sigmafc{}        & = \sigmafleftbottomc \text{ on } \pomleftbottomeqc,
\end{align}
where $\sigmafleftbottomc$ is chosen such that
\be\label{bc_fl_membrane_7_current}
\rhof \left[ \int \dint{\xcoord^1}  \wmem \left| \der{\bxc}{\xcoord^1}\right| + \int_{\pombottomeqc} \dint{\lc} \, \vfbottom \cdot \bnormalc \right] = 0,
\ee
and $\dint{\lc}$ id the boundary length-element in the \ac{cur} configuration \cite{kamenskyLectureNotesMAE}.

\Cref{bc_fl_membrane_1_current} implies that the fluid is injected at constant velocity from the bottom edge \pombottomc.  \Cref{bc_fl_membrane_2_current} is a no-slip \ac{bc} on the left edge \pominc, and \cref{bc_fl_membrane_3_current,bc_fl_membrane_4_current} follow from axial symmetry, cf. for example \cref{bc_fl_mem_dom_2,bc_fl_mem_dom_3}.
\Cref{bc_fl_membrane_5_current} implies that the motion of the fluid is geometrically conforming to that of the membrane   \cite{kamenskyLectureNotesMAE}, while it allows for a slip between the tangential flow of membrane and fluid material-elements---see \cref{sec_fl_mem_mem}. Also, in \cref{bc_fl_membrane_5_current}, $\bvfc{}$ is evaluated at a \ac{cur} location $\bxc$, and $\budomdot$ at the corresponding \ac{ref} location $\bxr$ related to $\bxc$ by $\budom$ through \cref{eq_def_u}.

Finally, we discuss \cref{bc_fl_membrane_6_current}. Let us consider the overall \ac{bvp} for the three sectors---membrane, fluid domain and fluid---given by \cref{\eqsmembrane,\eqsdom,\eqsbfluc} and the \acp{bc}, \cref{\bcsmembrane,\bcsdom,{bc_fl_membrane_1_current,bc_fl_membrane_2_current,bc_fl_membrane_3_current,bc_fl_membrane_4_current,bc_fl_membrane_5_current}}. Let us consider a translation of the fluid pressure field
\be \label{eq_translation_sigma}
\sigmafc{}{} \rightarrow \sigmafc{} + C,
\ee
where $C$ is independent of $\bxc$. The only components of the \ac{bvp} which break its invariance with respect to \cref{eq_translation_sigma} are the \acp{rhs} of \cref{eq-dyn-v,eq-dyn-w}, where $f_{\fluid\, \text{t}}$ and $\fflun$ depend on the absolute value of $\sigmafc{}$, see \cref{def_force_tan,def_force_norm,def_f_fluid_mem,eq_strees_tensor_fluid}. All the remaining equations in the \ac{bvp} depend on $\sigmafc{}$ through its derivatives only, and are thus invariant with respect to  \cref{eq_translation_sigma}.
By pinning the fluid pressure to the point $\pomleftbottomeqc$, we  thus tune the force exerted by the fluid on the membrane. We chose the pinned value, $\sigmafleftbottomc$, in such a way that the amount of fluid matter  per unit time which flows into $\omc$ through \pombottomc{} is equal to the amount of fluid matter per unit time which flows through \pomtopc. This condition is expressed by \cref{bc_fl_membrane_7_current}, and it physically reflects of the known role of pressure as Lagrange multiplier enforcing mass conservation.

Overall, we will thus consider \cref{\bcbflu} as \acp{bc} for the fluid.

\paragraph{Equations of motion in  \acl{ref} configuration coordinates}
\label{sec_fl_mem_flu_ref}

We will now rewrite the equations of motion and the \acp{bc} of \cref{sec_fl_mem_flu_cur} in terms of $\bvfr{}$, $\sigmafr{}$, and the \ac{ref} coordinate $\bxr$. The equations of motion \crefs{\eqsbfluc} yield, respectively,

\bw
\begin{align}
    \label{eq_fl_rigid_fl_1_reference}
    \rhof \left\{ \pder{ \vfr{\alpha}(\bxr, t)}{t} + \left[ \vfr{\gamma}(\bxr, t)  - \der{\varphi^{\gamma\, t}(\bxr)}{t}\right]  G_{\beta \gamma}(\budom^t(\bxr)) \frac{\partial  \vfr{\alpha}(\bxr, t)}{\partial \xr[\beta]} \right\} & =
    G_{\gamma \beta}(\budom^t(\bxr))  \pderr{\sigmastressr{\alpha}{\beta}}{\gamma},                                                                                                                                                           \\
    \label{eq_fl_rigid_fl_2_reference}
    G_{\beta \alpha}(\budom^t(\bxr)) \pderr{ \vfr{\alpha}(\bxr, t)}{\beta}                                                                                                                                                             & = 0,
\end{align}
\ew

which hold in $\omr$. In \cref{eq_fl_rigid_fl_1_reference,eq_fl_rigid_fl_2_reference} the tensor $G$ is given by the inverse of $F$ \cite{kamenskyLectureNotesMAE}:
\be
\label{eq_def_G}
G_{\alpha \beta}(\bu) \equiv [F(\bu)]^{-1}_{\alpha \beta}.
\ee\vspace{5mm}

\subsubsection{Temporal discretization}\label{sec_temp_discretization_mem}

To numerically solve for the dynamics in a time span $0 \leq t \leq T$, we discretize time by setting $\deltat \equiv T/N$, $t_n \equiv n\, \deltat$, with $n = 0, 1, \cdots, N$.

We combine the \ac{cn} discretization method for \ac{ns} equations \cite{loggAutomatedSolutionDifferential2012,crankPracticalMethodNumerical1947} with the \ac{ale} formulation, by discretizing the dynamical equations for the membrane, fluid domain and  fluid.
The  dynamics of the system is obtained as follows.\\ \vspace{5mm}

Following the \ac{cn} temporal-discretization scheme \cite{crankPracticalMethodNumerical1947}, we define fields at staggered,  integer and semi-integer time steps  \cite{loggAutomatedSolutionDifferential2012}: For example, we set
\begin{align}
    \label{eq_time_step_1}
    \bvfr{n}(\bxr)          & \equiv \bvfr{}(\bxr, t_n)  ,                               \\
    \label{eq_time_step_2}
    \sigmafr{\nmhalf}(\bxr) & \equiv \sigmafr{}\left(\bxr, \frac{t_n+t_{n-1}}{2}\right),
\end{align}
and similarly for other fluid and membrane fields.

\paragraph{\Acl{flu}}\label{sec_temp_discretization_mem_bflu}

The discretization of the equations of motion for  the fluid proceeds along the same lines as in \cref{sec_fl_rigid}, and it is split into three steps according to the \ac{ipcs} scheme \cite{loggAutomatedSolutionDifferential2012,chorinNumericalSolutionNavierStokes1968}: Approximated velocity, pressure correction and velocity correction, see \cref{sec_splitting_scheme_fl_rig_fl}. We will show only the main steps of the \ac{ipcs} scheme here, see \cref{sec_fl_rigid,sec_temp_discretization_mem_mem} for details.

The \acp{pde} for $\bvbarf$ and $\sigmafr{}$ are
\bw
\begin{align}
    \label{eq_fl_rigid_fl_1_reference_disc_aux}
    \rhof \left\{  \frac{\vbarf^\alpha - \vfr{n-1,\, \alpha}}{\deltat} + \left[ \frac{3}{2} ( \vfr{n-1,\, \gamma}  -  \udomdot^{n-1,\, \gamma}) \, G_{\beta \gamma}(\budom^{n-1})  - \frac{1}{2} ( \vfr{n-2,\, \gamma}  -  \udomdot^{n-2,\, \gamma}) \, G_{\beta \gamma}(\budom^{n-2})  \right] \frac{\partial  \Vf^\alpha}{\partial \xr[\beta]}\right\} & =              \newlinenn
    G_{\gamma \beta}(\budom^{n-1})  \,          \pderr{}{\gamma} \sigmastressr{\alpha}{\beta}(\bVf, \sigmafr{\ast}, \bu^{n-1}),
    \\
    \label{eq_poiss_like_phi}
    \frac{\rhof}{\deltat} G_{\beta \alpha}(\budom^{n-1}) \pderr{\vbarf^\alpha }{\beta}                                                                                                                                                                                                                                                                   & = \newlinenn
    G_{\gamma \alpha}(\budom^{n-1}) \pderr{}{\gamma}\left[ G_{\beta \alpha}(\budom^{n-1})  \,          \pderr{ \phif   }{\beta } \right]                                                                                                                                                                                                                 & ,
\end{align}
\ew
where we have set
\begin{align}
    \label{eq_def_Vf}
    \bVf           & \equiv\frac{ \bvbarf + \bvfr{n-1}}{2},     \\
    \label{eq_def_sigmarast}
    \sigmafr{\ast} & \equiv \sigmafr{\nmthalf},                 \\
    \label{eq_def_phi_pressure}
    \phif          & \equiv \sigmafr{\ast} - \sigmafr{\nmhalf}.
\end{align}

The \acp{bc} for \cref{eq_fl_rigid_fl_1_reference_disc_aux,eq_poiss_like_phi} are \cite{loggAutomatedSolutionDifferential2012}
\begin{align}
    \label{bc_fl_membrane_1_current_disc}
    \bvbarf                                         & =  \vfbottom  \text{ on } \pombottomeqr, \\
    \label{bc_fl_membrane_2_current_disc}
    \bvbarf                                         & =  \bzero  \text{ on } \pomineqr,        \\
    \label{bc_fl_membrane_3_current_disc}
    \vbarf^1                                        & =  0 \; \text{ on } \pomouteqr ,         \\
    \label{bc_fl_membrane_4_current_disc}
    G(\budom^{n-1})_{\alpha 1}\pderr{\Vf^2}{\alpha} & =  0 \; \text{ on } \pomouteqr ,
\end{align}
\begin{align}
    \label{bc_fl_membrane_5_current_disc}
    \bvbarf                                                   & = \budomdot^{n-1}  \text{ on } \pomtopeqr,        \\
    \label{bc_fl_membrane_7_current_disc}
    \normalr{\gamma}  G_{\gamma \alpha}(\budom^{n-1})  G_{\beta \alpha}(\budom^{n-1})
    \pderr{ \phif   }{\beta }                                 & = 0\newlinenn
    \text{ on } \pombottomeqr \cup \pomtopeqr \cup \pomineqr, &                                                   \\
    \label{bc_fl_membrane_8_current_disc}
    G(\budom^{n-1})_{\alpha 1} \pderr{\phif}{\alpha}          & = 0 \text{ on } \pomouteqr,                       \\
    \label{bc_fl_membrane_9_current_disc}
    \sigmafc{}                                                & = \phifleftbottomc \text{ on } \pomleftbottomeqc,
\end{align}
where in \cref{bc_fl_membrane_9_current_disc}, $\phifleftbottomc$ is chosen so as to satisfy the mass-conservation relation
\be\label{bc_fl_membrane_7_reference}
\rhof \left[ \int \dint{\xcoord^1}  \wmem^n \left| \der{}{\xcoord^1}(\bxr + \bumem^\nmhalf )\right| + \int_{\pombottomeqr} \dint{\lr} \, \vfbottom \cdot \bnormalr \right] = 0,
\ee
and $\dint{\lr}$ is the boundary length-element in the \ac{ref} configuration \cite{kamenskyLectureNotesMAE}.
\Cref{bc_fl_membrane_7_reference} is \cref{bc_fl_membrane_7_current} expressed in terms of \ac{ref} quantities.

Finally, $\bvfr{n}$ is determined from $\bvbarf $ and $\sigmafr{\nmhalf}$ by means of the relation

\be
\label{eq_phi_1}
\rhof \frac{\vbarf^\alpha - \vfr{n, \, \alpha}}{\deltat}  = G_{\beta \alpha}(\budom^{n-1})         \pderr{ \phif   }{\beta } \text{ in } \omr.
\ee\vspace{1cm}\\

\paragraph{\Acl{mem}}\label{sec_temp_discretization_mem_mem}

We discretize \cref{eq-dyn-v,eq-dyn-w,eq-dyn-continuity,eq-dyn-dX,eq-dyn-dX1,eq-dyn-dX2} as follows

\bw
\begin{align}
    \label{eq_mem_flu_disc_3}
    \nab^\nmhalf_i \vmem{i} - 2 H^\nmhalf \wmem^n                                                                              =                                                                                                                             & 0,                        \\
    \label{eq_mem_flu_disc_1}
    \rhom \Big(  \frac{\vmem{n,\, i} - \vmem{n-1,\, i}}{\deltat} +   \frac{3}{2}\vmem{n-1,\, j}  \nab^{\nmhalf}_j \vmem{n-1,\, i} -   \frac{1}{2}\vmem{n-2,\, j}  \nab^{\nmhalf}_j \vmem{n-2,\, i} - 2 \vmem{n-1,\, j} \wmem^{n-1} \fnij{b}{\nmhalf}{i}{j} - & \newlinenn
    \wmem^{n-1} \nab^{\nmhalf, \, i} \wmem^{n-1} \Big)                                                                                                                                                                                                       & =              \newlinenn
    - \nab^\nmhalf_j \Pi^{ij} \left( \frac{\bvmem{n} + \bvmem{n-1}}{2}, \frac{\wmem^n + \wmem^{n-1}}{2},\sigmam{\nmhalf},\nu^\nmhalf,\psi^\nmhalf \right)+ \newlinenn
    f_{\fluid\, \text{t}}^i({\bsigmastressr}(\bvfr{n}, \sigmafr{\nmhalf}, \budom^{n-1}), \nu^\nmhalf, \psi^\nmhalf) ,                                                                                                                                        &
    \\
    \label{eq_mem_flu_disc_2}
    \rhom \left( \frac{\wmem^n - \wmem^{n-1}}{\deltat} + \vmem{n-1,\, i} \vmem{n-1,\, j} b^\nmhalf_{ji} + \frac{3}{2} \vmem{n-1, \, i}  \nab^\nmhalf_i \wmem^{n-1}  - \frac{1}{2} \vmem{n-2, \, i}  \nab^\nmhalf_i \wmem^{n-2} \right)                       & =\newlinenn
    f_\lapl(\sigmam{\nmhalf}, \nu^\nmhalf,\psi^\nmhalf) + f_\kap(\nu^\nmhalf,\psi^\nmhalf)  + \fetamemn     \left( \frac{\bvmem{n} + \bvmem{n-1}}{2}, \frac{\wmem^n + \wmem^{n-1}}{2}, \nu^\nmhalf, \psi^\nmhalf\right) + \newlinenn
    \fflun ({\bsigmastressr}(\bvfr{n}, \sigmafr{\nmhalf}, \budom^{n-1}), \nu^\nmhalf, \psi^\nmhalf),                                                                                                                                                         &\end{align}
\ew

\begin{align}
    \label{eq_mem_flu_disc_4}
    \frac {\bumem^\nmhalf-\bumem^\nmthreehalf}{\deltat}                                                                                 = & \wmem^{n} \, \normalc{\nmhalf}, \\
    \label{eq_mem_flu_disc_5}
    \pder {\umem^{\nmhalf, \, 1}}{x^1}                                                                                            =       &
    - \pder{\xmemr{1}}{\xcoord^1} + \nu^\nmhalf \cos \psi^\nmhalf,                                                                                                          \\
    \label{eq_mem_flu_disc_6}
    \pder {\umem^{\nmhalf, \, 2}}{x^1}                                                                                               =    & - \pder{\xmemr{2}}{\xcoord^1}
    -\nu^\nmhalf \sin \psi^\nmhalf.
\end{align}

In \cref{eq_mem_flu_disc_1,eq_mem_flu_disc_2}, $\bsigmastressr$ denotes the fluid stress tensor in the \ac{ref} configuration:
\begin{align}
    \label{eq_def_sigma_ref}
    \sigmastressr{\alpha}{\beta}^t(\bxr)  \equiv & \sigmastressc{\alpha}{\beta}({\bm \varphi}^t(\bxr)) \newlinenn
    =                                            & \sigmafr{}(\bxr, t) \delta_{\alpha \beta} + \etaf \Big[ G_{\gamma \beta}(\budom^t(\bxr)) \pderr{\vfr{\alpha}}{\gamma} + \newlinenn
                                                 & G_{\gamma \alpha}(\budom^t(\bxr)) \pderr{\vfr{\beta}}{\gamma}\Big].
\end{align}

Proceeding along the same lines, the discrete forms of the \acp{bc} \crefs{\bcsmembrane} read, respectively,
\bw
\begin{align}
    \label{bc_fl_mem_mem_1_disc_bis}
    \normalmem^\nmhalf_j \Pi^{ij}\left( \frac{\bvmem{n} + \bvmem{n-1}}{2},\frac{\wmem^n + \wmem^{n-1}}{2}, \sigmam{\nmhalf}, \nu^\nmhalf, \psi^\nmhalf \right) & = 0 \text{ on } \pomlineineqr        \\
    \label{bc_fl_mem_mem_2_disc_bis}
    \vmem{n, \, i}                                                                                                                                             & = 0 \text{ on } \pomlineouteqr,      \\
    \label{bc_fl_mem_mem_3_disc_bis}
    \wmem^n                                                                                                                                                    & = 0 \text{ on } \pomlineineqr,       \\
    \label{bc_fl_mem_mem_4_disc_bis}
    \nab^\nmhalf_i \wmem^n                                                                                                                                     & = 0 \text{ on } \pomlineouteqr,      \\
    \label{bc_fl_mem_mem_5_disc_bis}
    \sigmam{\nmhalf}                                                                                                                                           & = 0 \text{ on } \pomlineineqr,       \\
    \label{bc_fl_mem_mem_6_disc_bis}
    \bumem^\nmhalf                                                                                                                                             & = {\bm 0} \text{ on } \pomlineineqr, \\
    \label{bc_fl_mem_mem_7_disc_bis}
    \umem^{\nmhalf, \, 1}                                                                                                                                      & = 0 \text{ on } \pomlineouteqr,      \\
    \label{bc_fl_mem_mem_8_disc_bis}
    \pder{\umem^{\nmhalf, \, 2}}{\xcoord^1}                                                                                                                    & = 0 \text{ on } \pomlineouteqr.
\end{align}
\ew

Following the \ac{ipcs} scheme,  in what follows we will split the \ac{bvp} above into three steps  \cite{loggAutomatedSolutionDifferential2012,chorinNumericalSolutionNavierStokes1968}.
The numerical stability and convergence of this scheme for a membrane on a deforming manifold have been discussed in \cite{worthmullerIRENEFluIdLayeR2025}, while we refer to \cite{formaggia1999a,boffi2004stability,sanmartinConvergenceFiniteElement2009} for a general stability and convergence analysis of \ac{ale} schemes for model problems.

\begin{enumerate}[label=\currentsec.\arabic*]
    \titleditem{Approximated velocities}{
        \label{item_mem_split_1}
        Proceeding along the lines of the splitting scheme for  the fluid, we introduce the auxiliary  velocities $\bvbarmem$ and $\wbarmem$, which approximate, respectively, $\bvmem{n}$ and $\wmem^n$ \cite{chorinNumericalSolutionNavierStokes1968}. We define $\bvbarmem$ and $\wbarmem$ as solutions of the \acp{pde}

        \bw
        \begin{align}
            \label{eq_fl_mem_mem_disc_1}
            \rhom \Big[  \frac{\vbarmem^i - \vmem{n-1,\, i}}{\deltat} +  \left( \frac{3}{2}\vmem{n-1,\, j} - \frac{1}{2} \vmem{n-2,\, j} \right)     \nab^{\nmhalf}_j \Vmem^i - 2 \Vmem^j \Wmem \fnij{b}{\nmhalf}{i}{j} -    \Wmem \nab^{\nmhalf, \, i} \Wmem \Big] & =              \newlinenn
            - \nab^\nmhalf_j \Pi^{ij}(\bVmem, \Wmem,\sigmamast,\nu^\nmhalf,\psi^\nmhalf) + \newlinenn
            f_{\fluid\, \text{t}}^i({\bsigmastressr}(\bvfr{n}, \sigmafr{\nmhalf}, \budom^{n-1}), \nu^\nmhalf, \psi^\nmhalf) ,
                                                                                                                                                                                                                                                                    &                           \\
            \label{eq_fl_mem_mem_disc_2}
            \rhom \left[ \frac{\wbarmem - \wmem^{n-1}}{\deltat} + \Vmem^i \Vmem^j b^\nmhalf_{ji} + \left( \frac{3}{2} \vmem{n-1, \, i} - \frac{1}{2}\vmem{n-2,\,i}\right) \nab^\nmhalf_i \Wmem \right]                                                              & =\newlinenn
            f_\lapl(\sigmamast, \nu^\nmhalf,\psi^\nmhalf) + f_\kap(\nu^\nmhalf,\psi^\nmhalf)  + \fetamemn      (\bVmem, \Wmem, \nu^\nmhalf, \psi^\nmhalf) + \newlinenn
            \fflun ({\bsigmastressr}(\bvfr{n}, \sigmafr{\nmhalf}, \budom^{n-1}), \nu^\nmhalf, \psi^\nmhalf)                                                                                                                                                         & ,
        \end{align}
        \ew
        where we have set
        \begin{align}
            \label{eq_def_Vmem}
            \Vmem^i    & \equiv \frac{\vbarmem^i + \vmem{n-1,\,i}}{2}, \\
            \label{eq_def_Wmem}
            \Wmem      & \equiv \frac{\wbarmem + \wmem^{n-1}}{2},      \\
            \label{eq_def_sigmamasst}
            \sigmamast & \equiv \sigmam{\nmthalf},
        \end{align}
        \Cref{eq_fl_mem_mem_disc_1,eq_fl_mem_mem_disc_2} are the analog of \cref{eq_mem_flu_disc_1,eq_mem_flu_disc_2}, respectively,  with a fixed pressure field $\sigmamast$ obtained from the preceding step. In what follows, we will synonymously use the terms `pressure' and `tension'  for conciseness \cite{worthmullerIRENEFluIdLayeR2025}.

        Overall, \cref{eq_fl_mem_mem_disc_1,eq_fl_mem_mem_disc_2} determine $\vbarmem$ and $\wbarmem$.

    }
    \titleditem{Pressure correction}{
        \label{item_mem_split_2}
        Subtracting \cref{eq_mem_flu_disc_1,eq_fl_mem_mem_disc_1} and neglecting $\odeltat$ \cite{worthmullerIRENEFluIdLayeR2025}, we obtain
        \be
        \label{eq_fl_mem_mem_disc_4}
        \rhom \frac{\vmem{n, \, i} - \vbarmem^i }{\deltat}                         =   - \nab^{\nmhalf, \, i}\phimem,
        \ee
        where we have defined the pressure increment as
        \begin{align}
            \label{eq_def_phimem}
            \phimem & \equiv \sigmamast - \sigmam{\nmhalf}.
        \end{align}
        Proceeding along the same lines for \cref{eq_mem_flu_disc_2,eq_fl_mem_mem_disc_2}, we obtain
        \be
        \label{eq_fl_mem_mem_disc_5}
        \wmem^n                                                                     =  \wbarmem.
        \ee
        By taking the covariant derivative of \cref{eq_fl_mem_mem_disc_4} and using the mass-conservation equation \crefs{eq_mem_flu_disc_3} and \cref{eq_fl_mem_mem_disc_5}, we obtain
        \begin{align}
            \label{eq_poisson_like_phi_mem}
            \nab^\nmhalf_i \nab^{\nmhalf,\, i} \phimem                                & =  \newlinenn
            \frac{\rhom}{\deltat} (\nab^\nmhalf_i \vbarmem^i - 2 H^\nmhalf \wbarmem). &
        \end{align}
        \Cref{eq_poisson_like_phi_mem} is a second-order \ac{pde} which determines $\phimem$, and thus the tension $\sigmam{\nmhalf}$ through \cref{eq_def_phimem}.
    }

    \titleditem{Velocity correction}{
        \label{item_mem_split_3}
        The membrane velocity $\bvmem{n}$ is determined by means of \cref{eq_fl_mem_mem_disc_4} in terms of $\bvbarmem$ and $\phimem$.
    }
\end{enumerate}

The \acp{bc} for \cref{eq_fl_mem_mem_disc_1,eq_fl_mem_mem_disc_2,eq_fl_mem_mem_disc_4,eq_fl_mem_mem_disc_5,eq_mem_flu_disc_4,eq_mem_flu_disc_5,eq_mem_flu_disc_6,eq_poisson_like_phi_mem} are
\bw
\begin{align}
    \label{bc_fl_mem_mem_1_disc}
    \normalmem^\nmhalf_j \Pi^{ij}( \bVmem,\Wmem, \sigmamast, \nu^\nmhalf, \psi^\nmhalf ) & = 0 \text{ on } \pomlineineqr              \\
    \label{bc_fl_mem_mem_2_disc}
    \vbarmem^i                                                                           & = 0 \text{ on } \pomlineouteqr,            \\
    \label{bc_fl_mem_mem_3_disc}
    \wbarmem                                                                             & = 0 \text{ on } \pomlineineqr,             \\
    \label{bc_fl_mem_mem_4_disc}
    \nab^\nmhalf_i \wbarmem                                                              & = 0 \text{ on } \pomlineouteqr,            \\
    \label{bc_fl_mem_mem_5_disc}
    \phimem                                                                              & = 0 \text{ on } \pomlineineqr,             \\
    \label{bc_fl_mem_mem_6_disc}
    \normalmem^\nmhalf_i  \nab^{\nmhalf, \, i} \phimem                                   & = 0 \text{ on } \pomlineouteqr, \newlinenn
    \crefineq{bc_fl_mem_mem_6_disc_bis,bc_fl_mem_mem_7_disc_bis,bc_fl_mem_mem_8_disc_bis}.
\end{align}
\ew

In \cref{bc_fl_mem_mem_6_disc}, $\bm \normalmem$ is the unit normal to \pomline{}: It lies in the tangent bundle of $\omlineeq$, it is directed  outside $\omlineeq$, and it is normalized according to $\normalmem^i \normalmem_i = 1$ \cite{reallGeneralRelativity,marchiafavaAppuntiDiGeometria2005,worthmullerIRENEFluIdLayeR2025}.

\paragraph{\Acl{dom}}\label{sec_temp_discretization_mem_dom}

The discretization of the equations of motion for  the fluid domain proceeds along the same lines as in \cref{sec_fl_rigid}---see \cref{sec_splitting_scheme_fl_rig_dom} for details.
The discrete version of the \acp{pde} \crefs{\eqsdom} is

\begin{align}
    \label{eq_domain_fluid_rigid_disc}
    \pder{\Sdom_{\alpha \beta}(\budom^n)}{\xr[\beta]}          & = 0 \text{ in } \omr, \\
    \label{eq_domain_fluid_rigid_disc_dot}
    \pder{}{\xr[\beta]}\der{\Sdom_{\alpha \beta}(\budom^n)}{t} & = 0 \text{ in } \omr,
\end{align}
with \acp{bc} given by the discrete version of \cref{\bcsdom}:
\begin{align}
    \label{bc_fl_mem_dom_1_disc}
    \budom^n                          & = \bzero \text{ on } \pomineqr \cup \pombottomeqr,     \\
    \label{bc_fl_mem_dom_2_disc}
    \udom^{n, \, 1}                   & = 0 \text{ on } \pomouteqr,                            \\
    \label{bc_fl_mem_dom_3_disc}
    \pder{\udom^{n, \, 2}}{\xr[1]}    & = 0 \text{ on } \pomouteqr,                            \\
    \label{bc_fl_mem_dom_4_disc}
    \budom^n                          & = \bumem^{\nmhalf} \text{ on } \pomtopeqr,             \\
    \label{bc_fl_mem_dom_5_disc}
    \budomdot^n                       & = \bzero \text{ on } \pomineqr \cup \pombottomeqr,     \\
    \label{bc_fl_mem_dom_6_disc}
    \udomdot^{n, \, 1}                & = 0 \text{ on } \pomouteqr,                            \\
    \label{bc_fl_mem_dom_7_disc}
    \pder{\udomdot^{n, \, 2}}{\xr[1]} & = 0 \text{ on } \pomouteqr,                            \\
    \label{bc_fl_mem_dom_8_disc}
    \budomdot^n                       & = \wmem^n \, \normalc{\nmhalf} \text{ on } \pomtopeqr.
\end{align}
In \cref{bc_fl_mem_dom_4_disc,bc_fl_mem_dom_8_disc} the \ac{lhs} is evaluated at a \ac{ref} coordinate $\bxr$, and the \ac{rhs} at the corresponding membrane curvilinear coordinate $\xcoord^1$ such that $\bxmemr(\xcoord^1) = \bxr$.

Given the membrane fields $\vmem{n-1, i}$, $\wmem^{n-1, \, i}$, $\sigmam{\nmthalf}$, $\bumem^{\nmthalf}$, $\nu^{\nmthalf}$, $\psi^{\nmthalf}$, the fluid-domain fields $\budom^{n-1}$, $\budomdot^{n-1}$, and the fluid fields $\bvfr{n-1}$, $\sigmafr{\nmthalf}$, we determine the fields at the next time step $n \rightarrow n+1$ as follows:

\begin{enumerate}
    \titleditem{Update the fluid}{ \label{fl_mem_step_3}
        We determine the fluid fields $\bvmem{n} $ and $\sigmafr{\nmhalf}$ by solving the \ac{bvp} given by  \cref{eq_fl_rigid_fl_1_reference_disc_aux,eq_poiss_like_phi,eq_phi_1,bc_fl_membrane_1_current_disc,bc_fl_membrane_2_current_disc,bc_fl_membrane_3_current_disc,bc_fl_membrane_4_current_disc,bc_fl_membrane_5_current_disc,bc_fl_membrane_7_current_disc,bc_fl_membrane_8_current_disc,bc_fl_membrane_9_current_disc,bc_fl_membrane_7_reference},
    }

    \titleditem{Update the membrane}{ \label{fl_mem_step_1}
        The membrane fields $\vmem{n, i}$, $\wmem^{n, \, i}$, $\sigmam{\nmhalf}$, $\bumem^{\nmhalf}$, $\nu^{\nmhalf}$, $\psi^{\nmhalf}$ are determined by solving the \ac{bvp} given by
        \cref{eq_fl_mem_mem_disc_1,eq_fl_mem_mem_disc_2,eq_fl_mem_mem_disc_4,eq_fl_mem_mem_disc_5,eq_mem_flu_disc_4,eq_mem_flu_disc_5,eq_mem_flu_disc_6,eq_poisson_like_phi_mem,bc_fl_mem_mem_1_disc,bc_fl_mem_mem_2_disc,bc_fl_mem_mem_3_disc,bc_fl_mem_mem_4_disc,bc_fl_mem_mem_5_disc,bc_fl_mem_mem_6_disc,bc_fl_mem_mem_6_disc_bis,bc_fl_mem_mem_7_disc_bis,bc_fl_mem_mem_8_disc_bis}.
    }

    \titleditem{Update the fluid domain}{ \label{fl_mem_step_2}
        We determine the domain displacement field and its time derivative, $\budom^n$ and $\budomdot^n$, from the \acp{bvp} given by \cref{eq_domain_fluid_rigid_disc,eq_domain_fluid_rigid_disc_dot,bc_fl_mem_dom_1_disc,bc_fl_mem_dom_2_disc,bc_fl_mem_dom_3_disc,bc_fl_mem_dom_4_disc,bc_fl_mem_dom_5_disc,bc_fl_mem_dom_6_disc,bc_fl_mem_dom_7_disc,bc_fl_mem_dom_8_disc}.

    }

\end{enumerate}

The fields $\bvfc{n}$, $\sigmafc{\nmhalf}$ in the \ac{cur} configuration are then obtained from the respective fields in the \ac{ref} configuration through \cref{eq_vf_ref,eq_sigmaf_ref}.\\ \vspace{5mm}

An example of this dynamics is shown in \cref{fig_fluid_mem_mem,fig_fluid_mem_bflu}, where we show the interaction between air (the fluid) and an elastomer (the membrane).
The  channel has dimensions $L = 1 \, \met$, $h = 0.2 \, \met$, and  we regularly remesh the \acl{dom} in order to keep mesh deformations small enough \cite{doneaChapter14Arbitrary2004}. Proceeding along the lines of \cref{sec_fl_rigid},   membrane parameters have been estimated by considering a three-dimensional material, multiplying its parameters by a unit length corresponding to the out-of-plane depth, and by the membrane thickness, chosen to be $\delta = 10^{-3}\, \met$.

The membrane bending rigidity $\kaptwod$ for a two-dimensional plate  has been estimated from the relation $\kaptwod = E \delta^3/[12(1-\poissonratio^2)]$, which expresses $\kaptwod$ as a function of the Young modulus $E$, the Poisson ratio $\poissonratio$, and  $\delta$ \cite{landauTheoryElasticity1986}.
Typical values for an elastomer   are $E \sim 10^7 \, \pas$ \cite{fengPreparationCharacterizationSilicone2017} and $\poissonratio \sim 0.5$ \cite{mullerQuickAccurateMethod2019}, yielding $\kaptwod = 10^{-3} \, \kg \, \met^2/\second^2$, and thus $\kap =  10^{-3} \, \kg \, \met^3/\second^2$.
We chose $\rhom = 1 \, \kg /  \met$ and $\etam = 1 \, \kg\, \met /\second $ as the density \cite{caiSoftPolydimethylsiloxaneElastomers2015} and viscosity \cite{pochivalovDevelopmentVibrationDamping2021} of an elastomer, respectively.
The fluid parameters have been estimated from the air parameters:  $\rhof = 1 \, \kg/\met^2$, $\etaf = 10^{-5} \, \kg / \second$ \cite{hilsenrath1955tables}.
Also, we take $\vfbottom^1 =0$, and $\vfbottom^2$ corresponds to a Poiseuille flow \cite{landauFluidMechanics1987}, i.e.,  $\vfbottom^2(\bxc) =  \frac{3}{2 L^2}  \xc[1] (2 L -\xc[1]) v_0$, where $v_0 = 10^{-1}  \, \met / \second$.  Finally, the fluid-domain exponent is $\zeta = 3$ \cite{kamenskyLectureNotesMAE}.

\Cref{fig_fluid_mem_mem} demonstrates that our method allows for the formation of membrane overhangs---configurations in which the membrane cannot be represented as a height function $y$ of the abscissa $x$ \cite{worthmullerIRENEFluIdLayeR2025}. These configuration are allowed by the generalized \acl{al} parametrization \crefs{eq_intro_nu}, and they would not be accessible in other parametrizations, such as the Monge gauge \cite{hsiungFirstCourseDifferential1981}.

Finally, the dynamics of \cref{fig_fluid_mem_mem} corresponds to a fully inertial regime. In fact, the length scale to which the fluid flow is constrained is the radius of curvature $r_\mem \sim 1 \, \met$ of the membrane shape in the \ac{cur} configuration. Combined with the parameters of \cref{sec_fl_mem}, this yields a Reynolds number  $R \sim \rhof v_0 r_\mem / \etaf \sim 10^4$ \cite{landauFluidMechanics1987}.

\section{Discussion}
\label{sec_discussion}

\acresetall

We developed a  study on the interaction between a fluid and a \acl{mem}. Our analysis is based on the \ac{fe} method, and it leverages the \ac{fenics} \cite{loggAutomatedSolutionDifferential2012}, combined with the \ac{irene} \cite{worthmullerIRENEFluIdLayeR2025}.
In order to guide and introduce the reader to the interaction problem between a \acl{flu} and \acl{mem}, we provided two introductory examples of fluid-structure interaction \cite{kamenskyLectureNotesMAE}, in which the \acl{flu} interacts with a \acl{bod} and with an \acl{ela}, see \cref{sec_fl_rigid,sec_fl_elastic}, respectively. These Sections allowed to introduce the necessary physical an mathematical quantities to introduce the interaction between a fluid and a membrane.

In particular, the time-iteration scheme that we propose implies a partitioned resolution scheme \cite{felippaPartitionedAnalysisCoupled2001} well known in \ac{fsi} problems. In this regard, we denote  the rigid body, the elastic body or the membrane, by the `structure' with which the fluid interacts, in \cref{sec_fl_rigid,sec_fl_elastic,sec_fl_mem}. In such partitioned scheme one solves---for example---for the structure degrees of freedom in a given, fixed  configuration of the fluid and fluid domain. A similar partitioned strategy is used for the interaction between a fluid and a membrane---see \cref{fl_mem_step_3,fl_mem_step_1,fl_mem_step_2} in \cref{sec_temp_discretization_mem}.

Partitioned solutions are known to suffer for numerical instabilities \cite{forsterArtificialAddedMass2007}. Qualitatively speaking, as the structure movement displaces
the fluid elements, for the structure to accelerate with an acceleration $\bm a $,
an effective `added' fluid mass, $\madded$, needs to be displaced.
This mass results into an effective force  $\sim \bm{a}\,  \madded$ that the fluid exerts on the structure.
The partitioned scheme turns out to be stable if such force is significantly smaller than the structure inertia, ${\bm a} \, \mstr$ \cite{causinAddedmassEffectDesign2005}, i.e., if
\be\label{eq_added_mass}
\frac{\madded}{\mstr}\ll 1.
\ee

First, for the motion of a rigid body in a fluid of \cref{sec_fl_rigid}, $\madded \sim \rhof A_\structure \sim 2.3 \times 10^{-2}\, \kg$, and $\mstr = \rhob A_\structure = 1.4 \times 10^{-1} \, \kg $, where $A_\structure = \pi a b$ is the area of the body. The stability condition \crefs{eq_added_mass} is thus satisfied  with an accuracy of $\madded/\mstr \sim 10^{-2}$.
Second, in the analysis of  \cref{sec_fl_elastic}, where the elastic body moves in a fluid, $\madded \sim \rhof A_\structure \sim 6.2 \times 10^{-2}\, \kg$, and $\mstr = \rhoela A_\structure = \parameter[6.2][1]{\kg}$, where $A_\structure \sim \pi (a b - r^2)$ is the area of the elastic body. The stability condition \crefs{eq_added_mass} is thus satisfied with an accuracy of $\madded/\mstr \sim 0.1$.

Third, we discuss  the added-mass instability for the interaction between a fluid and a Helfrich membrane of \cref{sec_fl_mem}. At the  time shown in  \cref{fig_fluid_mem_mem}, we have $\mstr \sim \rhom l \sim 1 \, \kg$, $\madded \sim \rhof  l^2 \sim 1 \, \kg$, where $l \sim 1 \, \met$ is the average  membrane length in the \acl{cur} configuration.
As a result,  \cref{eq_added_mass} is  border line; added and structure mass are comparable,   $\madded/\mstr \sim 1$. More importantly,   the \aclp{bc} used in \cref{sec_fl_mem_mem} allow for  membrane elements to freely flow in through the \acl{le} extremity: as a result,  $l$ grows in time.
Given that  $\mstr$ and $\madded$ scale with $l$ as $\sim l$ and $\sim l^2$, respectively, the longer the system evolves, the more the condition \crefs{eq_added_mass} will be violated. Overcoming this instability is a natural direction for future studies---see below.

\begin{figure*}[t]
    \centering
    \includegraphics[width=\textwidth]{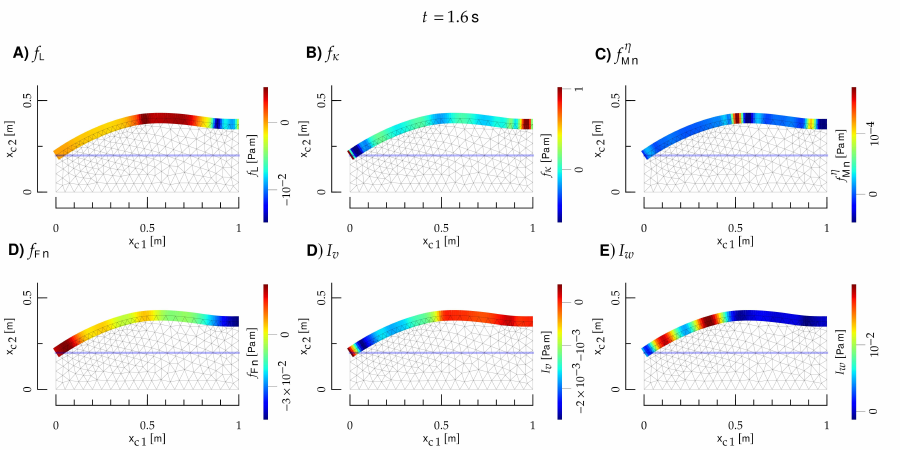}
    \caption{
        \label{fig_force_mem}
        Forces exerted on the \acf{mem} and membrane inertial terms, for the dynamics of \cref{fig_fluid_mem_mem,fig_fluid_mem_bflu}. The Figure follows the same notation as \cref{fig_fluid_mem_mem}. \plab{A}-\plab{D}: Laplace, elastic and viscous force, and force exerted on the membrane by the \acl{flu}, respectively, see \cref{eq_def_f_lap,eq_def_f_kap,eq_def_f_eta_n,def_force_norm}.
        \plab{E} and \plab{F}: Inertial membrane terms, see \cref{eq_def_I_v,eq_def_I_w}.
    }
\end{figure*}

Finally, our model allows for understanding  the physics of the bulge formation in \cref{fig_fluid_mem_mem,fig_fluid_mem_bflu}.
The \acp{bc} \crefs{\bcsmembrane} are such that the \acl{le} extremity of the  membrane, $\pomlineineqr$, is pinned, but its tangent angle $\psi$ is free. As a result, the  membrane tilts upwards under the force exerted on it by the fluid in the vicinity of $\pomlineineqr$, taking advantage of such free angle. The membrane thus bends in the vicinity of the symmetry axis $\pomlineouteqr$, increasing its local curcature, and thus accumulating an elastic stress.
The normal acceleration $\partial \wmem/\partial t$, which is referred to the outwards direction of the current normal $\normalc{}$ shown in  \cref{fig_ref_cur_mem}, receives contributions from multiple terms, see \cref{eq-dyn-w}. These are the forces, which appear in the \acl{rhs} of \cref{eq-dyn-w}, and inertial terms in the \acl{lhs}. First, the forces are the Laplace force $f_\lapl$, \cref{eq_def_f_lap}, the elastic force $f_\kap$, \cref{eq_def_f_kap}, the viscous force due to the flow of membrane elements $\fetamemn$ \cite{al-izziShearDrivenInstabilitiesMembrane2020}, \cref{eq_def_f_eta_n}, and the force $\fflun$ exerted by the fluid on the membrane, \cref{def_force_norm}. Second, the  inertial terms $I_v$, $I_w$ are given in \crefs{eq_def_I_v,eq_def_I_w}.
\Cref{fig_force_mem} depicts all such terms in an early stage of the bulge formation, showing that the elastic force $f_\kap$ dominates over all other terms. Such force releases the accumulated elastic stress above: It pushes the membrane downwards in the vicinity of $\pomlineineqr$ and upwards close to $\pomlineouteqr$---see \cref{fig_force_mem}B. Given that the membrane is pinned at $\pomlineineqr$, the force at $\pomlineineqr$ is neutralized by an equal and opposite force, cf. \cref{bc_fl_mem_mem_6}. On the other hand, at $\pomlineouteqr$, the membrane is free to move upwards driven by the elastic force, cf. \cref{bc_fl_mem_mem_7,bc_fl_mem_mem_7}, resulting into an increase of the membrane bulge.
As a result, the inflation mechanism is due to a combination of the free-angle \ac{bc} and elastic forces.

As a future direction of this work, we plan to go beyond the partitioned approach by using a monolithic scheme, where the degrees of freedom of the membrane,  fluid, and fluid domain are solved for at the same time \cite{heilEfficientSolverFully2004}. For a monolithic scheme, no added-mass instability exists, allowing for a characterization of a broader range of physical regimes, e.g.,  deformations larger  than those in \cref{fig_fluid_mem_mem,fig_fluid_mem_bflu}. This would allow for studying a large variety of physically sensible scenarios, e.g., the formation of  spikes in lipid membranes \cite{simonActinDynamicsDrive2019} and  membrane tubes \cite{derenyiFormationInteractionMembrane2002}.
Another natural future direction is the extension to three dimensions. Importantly, such extension would require a two-dimensional generalization of the \acl{al} gauge for the membrane, see \cref{sec_fl_mem_mem}. To achieve this, a natural parametrization would be the isothermal gauge \cite{carmoDifferentialGeometryCurves1976,worthmullerIRENEFluIdLayeR2025}. A time-dependent version of the isothermal gauge should then be worked out along the lines of the time-dependent \acl{al} gauge of \cref{sec_fl_mem_mem}.  This extension would allow for studying a large variety of additional phenomena, such non-axisymmetric, three-dimensional vesicle formation \cite{vasanMechanicalModelReveals2020}.

\section{List of Acronyms}\label{sec_acronyms}
\begingroup
\setlist[description]{itemsep=2pt, parsep=0pt, topsep=0pt}
\printacronyms[name=]
\endgroup

\begin{acknowledgments}
    We would like to thank S. Al-Izzi, S. Astruc and D. Lacoste for useful discussions, and  the \ac{fenics} online community for its precious support.

    This work was granted access to the high-performance computing resources of MesoPSL financed by the Region
    \^Ile de France, and to the Abacus cluster at Institut Curie.

    P. Sens and D. W\"orthm\"uller received funding from  European Research Council (ERC) grant ERC-SyG 101071793, awarded to P. Sens.

\end{acknowledgments}

\appendix

\acresetall

\numberwithin{equation}{section}
\numberwithin{figure}{section}
\section{Fluid and rigid body}\label{app_sec_fl_rigid}

\subsection{\Acl{bod} motion}\label{sec_rigid_body_body}

In this Section, we will work out the equations of motion for the rigid body.

\subsubsection{Equation of motion in  \acl{cur}-configuration coordinates}\label{sec_rigid_body_body_current}

The equation of motion for the rigid body, in \ac{cur} coordinates, reads \cite{goldsteinClassicalMechanics2004}

\begin{align}
        \label{eq_fl_rigid_rigid_current_1}
        I         \der{\omega}{t}
               & =                                                                                             - \int_\pomellipseeqc \dint{l_\current} \, \epsilon_{3 \alpha \beta} \, \sigmastressc{\beta}{\gamma} \normalc{\gamma} , \\
        \label{eq_fl_rigid_rigid_current_2}
        \omega & \equiv \der{\theta}{t}.
\end{align}
\Cref{eq_fl_rigid_rigid_current_1} relates the angular acceleration of the rigid body to the torque of external forces.
The moment of inertia of the body is denoted by $I$. The \ac{rhs} of \cref{eq_fl_rigid_rigid_current_1} contains the line integral, along the boundary of the body, of the torque of the local force exerted by  the fluid on  the body, with respect to the body left focal point $\bfocalpoint$ \cite{landauFluidMechanics1987}.
This force is written in terms of the fluid stress tensor $\bsigmastressc$ \cite{landauFluidMechanics1987}, expressed in \ac{cur} coordinates, which is given by \cref{eq_strees_tensor_fluid}, where $\etaf$ and $\rhof$ are the two-dimensional viscosity and density of the fluid, respectively.
Also, $\epsilon_{\alpha \beta \gamma}$ is the three-dimensional Levi-Civita symbol \cite{marchiafavaAppuntiDiGeometria2005}.
Finally,   $\dint{l_\current}$ is the line element of \pomellipsec in the \ac{cur} configuration, and $\bnormalc$ the unit boundary normal to \pomellipsec pointing outside $\omc$.

\subsubsection{Equation of motion in the \acl{ref}-configuration coordinates}\label{sec_rigid_body_body_reference}

In  \ac{ref} coordinates, \cref{eq_fl_rigid_rigid_current_1} reads

\bw
\begin{align}
        \label{eq_fl_rigid_rigid_reference}
        I \der{\omega}{t} & = \int_\pomellipseeqr \dint{s} \, \epsilon_{\alpha \beta} \left[ \varphi^{\alpha, \, t}(\bxi{s}) - \focalpoint^\alpha \right] \sigmastressr{\beta}{\gamma}(\bxi{s}) \epsilon_{\gamma \lambda} F_{\lambda \mu}(\budom^t(\bxi{s})) \der{\xi^\mu(s)}{s}.
\end{align}
\ew

The integral in the \ac{rhs} of  \cref{eq_fl_rigid_rigid_reference} is taken over the curvilinear boundary \pomellipser, which is parameterized as $\bxr = \bxi{s}$, where $s$ is the curvilinear coordinate. Also, $\bsigmastressr$ denotes the stress tensor in the \ac{ref} configuration, which is given by \cref{eq_def_sigma_ref}, where $G$ is given by \cref{eq_def_G}.

\subsection{Fluid motion}\label{sec_rigid_body_fl}

In this Section, we will work out the equations of motion for the fluid.

\subsubsection{Equations of motion in the \acl{cur}-configuration coordinates}\label{sec_rigid_body_fl_cur}

The equations of motion, expressed in terms of the \ac{cur} fields $\vfc$ and $\sigmafc{}$, are \cref{\eqsbfluc}, see \cref{sec_fl_mem_flu} for details.

The \acp{bc} for \cref{\eqsbfluc} are
\begin{align}
        \label{bc_fl_rigid_1_current}
        \vfc^1(\bxc)                  & =  \vfin \; \text{ on } \pomineqc,                                                                                   \\
        \label{bc_fl_rigid_2_current}
        \vfc^2  (\bxc)                & =  0 \; \text{ on } \pomineqc ,                                                                                      \\
        \label{bc_fl_rigid_3_current}
        \bvfc{}    (\bxc)             & = \bzero \; \text{ on } \pomweqc ,                                                                                   \\
        \label{bc_fl_rigid_4_current}
        \vfc^\alpha (\bxc)            & = \der{[R(\theta)]_{\alpha \beta} }{t} [\psi^{t \, \beta}(\bxc) - \focalpoint^\beta] \; \text{ on } \pomellipseeqc , \\
        \label{bc_fl_rigid_5_current}
        \etaf \pderc{ \vfc^\alpha}{1} & = 0  \; \text{ on } \pomouteqc ,                                                                                     \\
        \label{bc_fl_rigid_6_current}
        \sigmafc{} (\bxc)             & = 0 \; \text{ on } \pomouteqc.
\end{align}
\Cref{bc_fl_rigid_1_current,bc_fl_rigid_2_current} ensure that the fluid is injected on the boundary \pominc{} in a direction parallel to the $\xc[1]$ axis, and \cref{bc_fl_rigid_3_current} is a no-slip \ac{bc} \cite{landauFluidMechanics1987} along the channel walls \pomwc, see \cref{fig_ref_curr_rigid}.
\Cref{bc_fl_rigid_4_current} ensures that, at the boundary \pomellipsec{},  the fluid velocity matches the rotational velocity of the  body. Given that the  body can only rotate about its left focal point $\bfocalpoint$, the rotational velocity of a material  point $\bxc \in \pomellipseeqc$ of the body is given by
\be
\der{[R(\theta)]_{\alpha \beta}}{t}[\psi^{t \, \beta}(\bxc) - c_\beta],
\ee
where $\bm R$ is the  rotation matrix corresponding to a counterclockwise rotation:
\be
{\bm R}(\theta) \equiv
\left(
\begin{matrix}
                \cos \theta & - \sin \theta \\
                \sin \theta & \cos \theta
        \end{matrix}
\right).
\ee
Finally, \cref{bc_fl_rigid_5_current,bc_fl_rigid_6_current} ensure, respectively, that traction and tension vanish on \pomoutc; this corresponds to a free-flow hypothesis at \pomoutc \cite{loggAutomatedSolutionDifferential2012,landauFluidMechanics1987}.

\subsubsection{Equations of motion in the \acl{ref}-configuration coordinates}\label{sec_rigid_body_fl_ref}

We will now rewrite the equations of motion and the \acp{bc}   in terms of $\bvfr{}$, $\sigmafr{}$, and the \ac{ref} coordinate $\bxr$. The equations of motion \crefs{\eqsbfluc} yield, respectively, \cref{eq_fl_rigid_fl_1_reference,eq_fl_rigid_fl_2_reference}, see \cref{sec_fl_mem_flu_ref}.
The \acp{bc}  \crefs{bc_fl_rigid_1_current,bc_fl_rigid_2_current,bc_fl_rigid_3_current,bc_fl_rigid_4_current,bc_fl_rigid_5_current,bc_fl_rigid_6_current} yield as

\begin{align}
        \label{bc_fl_rigid_1_reference}
        \vfr{1}(\bxr)                                                  & =  \vfin \; \text{ on } \pomineqr,                                                                      \\
        \label{bc_fl_rigid_2_reference}
        \vfr{2}      (\bxr)                                            & =  0 \; \text{ on } \pomineqr ,                                                                         \\
        \label{bc_fl_rigid_3_reference}
        {\bvfr{}} (\bxr)                                               & = {\bm 0} \; \text{ on } \pomweqr ,                                                                     \\
        \label{bc_fl_rigid_4_reference}
        \vfr{\alpha}(\bxr)                                             & = \der{[R(\theta)]_{\alpha \beta} }{t} (\xr[\beta] - \focalpoint^\beta) \; \text{ on } \pomellipseeqr , \\
        \label{bc_fl_rigid_5_reference}
        \etaf G_{\beta 1}(\budom^t(\bxr)) \pderr{ \vfr{\alpha}}{\beta} & = 0  \; \text{ on } \pomouteqr ,                                                                        \\
        \label{bc_fl_rigid_6_reference}
        \sigmafr{} (\bxr)                                              & = 0 \; \text{ on } \pomouteqr.
\end{align}

\subsection{Fluid-domain motion}
\label{sec_eq_fluid_domain}

\begin{figure*}[t]
        \centering
        \includegraphics[width=\textwidth]{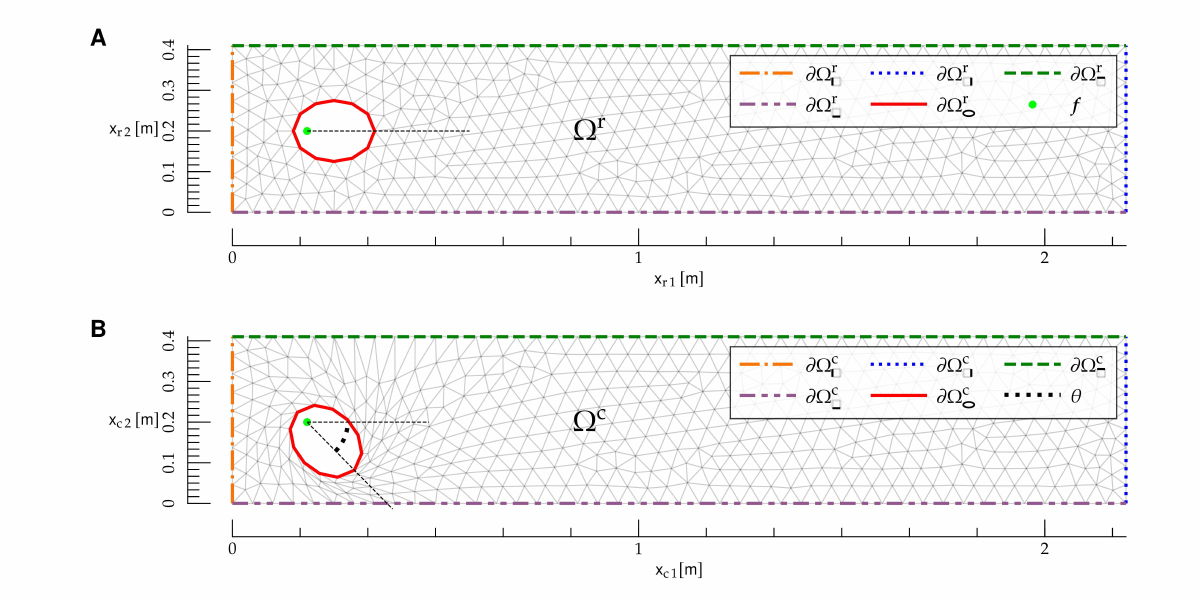}
        \caption{
                \label{fig_ref_curr_rigid}
                \Acf{ref} and \acf{cur} configuration for the interaction between a \acl{flu} and a \acl{bod}.
                \plab{A} \ac{ref} configuration. The region $\omr$ is given by the rectangle $0 \leq \xr[1] \leq L$, $0 \leq \xr[2] \leq h$ with the elliptical hole (the  body), and is delimited by the mesh (gray lines).  Boundaries in the \ac{ref} configuration are denoted by $\pomineqr$, $\pomouteqr$, $\pomtopeqr$, $\pombottomeqr$ and $\pomellipseeqr$ (colored dashed lines). The ellipse focal point $\bfocalpoint$ (light green dot)  is also shown.
                \plab{B} \ac{cur} configuration. The region $\omc$ is delimited by the mesh (gray lines). Boundaries in the \ac{cur} configuration are denoted by $\pomineqc$, $\pomouteqc$, $\pomtopeqc$, $\pombottomeqc$ and $\pomellipseeqc$ (colored dashed lines), and the rotational angle $\theta$ of the  body is shown as an arc (black dotted line). The Cartesian coordinates $\bxc$ are also shown.
        }
\end{figure*}

The equations of motion \crefs{\eqsfluidbodyref} involve the displacement field $\budom$, which describes how the region---or domain---which contains the fluid, deforms over time, cf. \cref{fig_ref_curr_rigid}. We will denote this region by the fluid domain.
We observe that this field has, so far, been left arbitrary, and we have the freedom to choose it.

Here, $\budom$ will be determined through an analogy with the theory of elasticity \cite{kamenskyLectureNotesMAE}. We imagine that, as the  body rotates about its focal point,  each point $\bxr \in \omr$ moves into a point $\bxc \in \omc$, and that the domain $\omr$ is deformed, e.g., locally stretched and compressed, into $\omc$, like an elastic medium. We will determine $\budom$ as the solution of a \ac{bvp} given by the  elastic model \crefs{eq_mesh_fluid_rigid} \cite{landauTheoryElasticity1986,kamenskyLectureNotesMAE}, see \cref{sec_fl_mem_dom} for details.

The \acp{bc} for the  \ac{pde} \crefs{eq_mesh_fluid_rigid} are
\begin{align}
        \label{bc_ela_1}
        \budom & =\bzero \text{ on } \pomsqeqr,                                                                 \\
        \label{bc_ela_2}
        \budom & = \bfocalpoint - \bxr + {\bm R}(\theta)\cdot (\bxr - \bfocalpoint) \text{ on } \pomellipseeqr.
\end{align}
\Cref{bc_ela_1} ensures that the mesh is not deformed at the boundary \pomsqr, while \cref{bc_ela_2} imposes that the mesh deformation at \pomellipser  matches the deformation induced by the rotation of the body.

\Cref{eq_mesh_fluid_rigid,bc_ela_1,bc_ela_2} constitute the \ac{bvp} which determines $\budom$. Along the lines of \acp{bvp} in the theory of elasticity, such \ac{bvp} is already, naturally formulated in the \ac{ref} configuration \cite{landauTheoryElasticity1986}.

From  the \ac{bvp} above, we obtain a \ac{bvp} for $\budomdot$, which will be needed in the following to describe the fluid-domain motion. By deriving both sides of \cref{eq_mesh_fluid_rigid,bc_ela_1,bc_ela_2}, we obtain, respectively, \cref{eq_mesh_fluid_rigid_dot} and
\begin{align}
        \label{bc_ela_1_dot}
        \budomdot & =\bzero \text{ on } \pomsqeqr,                                                         \\
        \label{bc_ela_2_dot}
        \budomdot & =  \omega \der{{\bm R}}{\theta}\cdot (\bxr - \bfocalpoint) \text{ on } \pomellipseeqr,
\end{align}
where in \cref{bc_ela_1_dot,bc_ela_2_dot} we used the fact that $\bxr$ is independent of time, and we substituted  \cref{eq_fl_rigid_rigid_current_2} in \cref{bc_ela_2_dot}. The \ac{lhs} of \cref{eq_mesh_fluid_rigid_dot} depends on the second partial derivatives of $\budom$ and $\budomdot$---see \cref{eq_def_S,eq_ela_1,eq_ela_2,eq_ela_3,eq_ela_4}. As a result, given $\theta$, $\omega$, and $\budom$ from \cref{eq_mesh_fluid_rigid,bc_ela_1,bc_ela_2}, the \ac{bvp} \crefs{eq_mesh_fluid_rigid_dot,bc_ela_1_dot,bc_ela_2_dot}  determines $\budomdot$.\vspace{0.5cm}\\

Overall \cref{eq_fl_rigid_rigid_current_2,eq_fl_rigid_rigid_reference,eq_fl_rigid_fl_1_reference,eq_fl_rigid_fl_2_reference,eq_mesh_fluid_rigid,eq_mesh_fluid_rigid,bc_fl_rigid_1_reference,bc_fl_rigid_2_reference,bc_fl_rigid_3_reference,bc_fl_rigid_4_reference,bc_fl_rigid_5_reference,bc_fl_rigid_6_reference,bc_ela_1,bc_ela_2,eq_mesh_fluid_rigid_dot,bc_ela_1_dot,bc_ela_2_dot}
constitute a \ac{bvp} for $\theta, \omega, \bvfr{}, \sigmafr{}$, $\budom$ and $\budomdot$ which, combined with the temporal \acp{bc}
\begin{align}
        \theta(t=0)             & = \theta_0              \\
        \omega(t=0)             & = \omega_0              \\
        {\bvfr{}}(\bxr, t=0)    & = {\bvfr{}}_0(\bxr),    \\
        {\sigmafr{}}(\bxr, t=0) & = {\sigmafr{}}_0(\bxr),
\end{align}
determine the dynamics of the system.

\subsection{Temporal discretization}\label{sec_temp_discretization_rig}

To numerically solve for the dynamics in a time span $0 \leq t \leq T$, we discretize time by setting $\deltat \equiv T/N$, $t_n \equiv n\, \deltat$, with $n = 0, 1, \cdots, N$.

We combine the \ac{cn} discretization scheme for \ac{ns} equations \cite{loggAutomatedSolutionDifferential2012,crankPracticalMethodNumerical1947} with the \ac{ale} formulation, by discretizing the dynamical equations for the rigid body, the fluid and the fluid domain---see below.

\subsubsection{\Acl{bod}}

Proceeding along the lines of the \ac{cn} scheme \cite{crankPracticalMethodNumerical1947}, we define fields at staggered,  integer and semi-integer time steps  \cite{loggAutomatedSolutionDifferential2012}---see for example \cref{eq_time_step_1,eq_time_step_2} for $\bvfr{}$ and $\sigmafr{}$---and similarly for other quantities.
We discretize \cref{eq_fl_rigid_rigid_reference,eq_fl_rigid_rigid_current_2} and, neglecting $\odeltat$, we obtain
\bw
\begin{align}
        \label{eq_def_omega_disc}
        \frac{\theta^n - \theta^{n-1}}{\deltat}     = & \omega^n       ,                                                                         \\
        \label{eq_fl_rigid_rigid_reference_disc}
        \frac{\omega^n - \omega^{n-1}}{\deltat}  =    &
        I  \int_\pomellipseeqr \dint{s} \, \epsilon_{\alpha \beta} \left[ \xi^\alpha(s) + u^{n-1,\, \alpha}(\bxi{s})- \focalpoint^\alpha \right]                                                                             \sigmastressr{\beta}{\gamma}^{n-1}(\bxi{s}; {\bvfr{n-1}, \sigmafr{\nmthalf}}) \times \newlinenn
                                                      & \epsilon_{\gamma \lambda} F_{\lambda \mu}(\budom^{n-1}(\bxi{s})) \der{\xi^\mu(s)} {s}  .
\end{align}
\ew
In \cref{eq_fl_rigid_rigid_reference_disc} we used \cref{eq_def_u,eq_def_phi} and we  explicitly indicated the dependence of the stress tensor \crefs{eq_def_sigma_ref} on $\bm \vfc$ and $\sigmafc{}$.

\subsubsection{\Acl{flu}}
\label{sec_splitting_scheme_fl_rig_fl}

To discretize in time the fluid problem, we  apply the \ac{cn} splitting scheme \cite{loggAutomatedSolutionDifferential2012,crankPracticalMethodNumerical1947} to the \ac{ns} equations \crefs{eq_fl_rigid_fl_1_reference,eq_fl_rigid_fl_2_reference} in the \ac{ref} configuration. This splitting scheme is derived form the \ac{ipcs} \cite{godaMultistepTechniqueImplicit1979}, in which the solution of the \ac{ns} equations is split into three intermediate steps. In what follows, we will synonymously use the terms `pressure' and `surface tension'  for conciseness \cite{worthmullerIRENEFluIdLayeR2025}.

The discrete form of the \acp{pde} \crefs{eq_fl_rigid_fl_1_reference,eq_fl_rigid_fl_2_reference} is
\bw
\begin{align}
        \label{eq_fl_rigid_fl_1_reference_disc}
        \rhof \Bigg[  \frac{\vfr{n, \, \alpha} - \vfr{n-1,\, \alpha}}{\deltat} + \frac{3}{2} ( \vfr{n-1,\, \gamma}  -  \dot{u}^{n-1,\, \gamma}) \, G_{\beta \gamma}(\budom^{n-1}) \frac{\partial  \vfr{n-1, \, \alpha}}{\partial \xr[\beta]} -
        \frac{1}{2} ( \vfr{n-2,\, \gamma}  -  \dot{u}^{n-2,\, \gamma}) \times                                                                                                                                                                                                                                                              & \newlinenn
        G_{\beta \gamma}(\budom^{n-2}) \frac{\partial  \vfr{n-2, \, \alpha}}{\partial \xr[\beta]} \Bigg]                                                                                                                                                                                                                                   & =              \newlinenn
        G_{\beta \alpha}(\budom^{n-1})  \,          \pderr{ \sigmafr{\nmhalf}   }{\beta }                + \etaf \,              G_{\delta \beta}(\budom^{n-1})            \,         \pderr{}{\delta} \left[  G_{\gamma \beta}(\budom^{n-1}) \pderr{ }{\gamma}\left(\frac{\vfr{n, \, \alpha} + \vfr{n-1, \, \alpha}}{2}\right)   \right], &                           \\
        \label{eq_fl_rigid_fl_2_reference_disc}
        G_{\beta \alpha}(\budom^{n-1}) \pderr{ \vfr{n,\, \alpha}}{\beta}                                                                                                                                                                                                                                                                   & = 0,
\end{align}
\ew
where we omitted the dependence on $\bxr$ for clarity. Also, in  the \ac{lhs} of \cref{eq_fl_rigid_fl_1_reference_disc} we rewrote the convective term  according to the Adams-Bashforth  discretization scheme \cite{SolvingOrdinaryDifferential1993,loggAutomatedSolutionDifferential2012}, and we used \cref{eq_def_u,eq_def_phi} to rewrite the time derivative of $\bm \varphi$ in terms of $\budomdot$.

The discrete form of the \acp{bc} \crefs{bc_fl_rigid_1_reference,bc_fl_rigid_2_reference,bc_fl_rigid_3_reference,bc_fl_rigid_4_reference,bc_fl_rigid_5_reference,bc_fl_rigid_6_reference} reads
\bw
\begin{align}
        \label{bc_fl_rigid_1_reference_disc}
        \vfr{n, \, 1}                                                                                                       & = \vfin \; \text{ on }\pomineqr,                                                                                                          \\
        \label{bc_fl_rigid_2_reference_disc}
        \vfr{n, \, 2}                                                                                                       & =  0 \; \text{ on } \pomineqr ,                                                                                                           \\
        \label{bc_fl_rigid_3_reference_disc}
        \bvfr{n}                                                                                                            & = {\bm 0} \; \text{ on } \pomweqr ,                                                                                                       \\
        \label{bc_fl_rigid_4_reference_disc}
        {\bvfr{}}                                                                                                           & =
        \omega^n  \left. \der{{\bm R}}{\theta} \right|_{\theta = \theta_n} \cdot (\bxr - \bfocalpoint)  \text{ on } \pomellipseeqr ,                                                                                                                                    \\
        \label{bc_fl_rigid_5_reference_disc}
        \etaf \, G_{\beta 1}(\budom^{n-1}) \pderr{}{\beta}\left( \frac{\vfr{n, \, \alpha} + \vfr{n-1, \, \alpha}}{2}\right) & = 0                                                                                                           \; \text{ on } \pomouteqr , \\
        \label{bc_fl_rigid_6_reference_disc}
        \sigmafr{\nmhalf}                                                                                                   & = 0 \; \text{ on } \pomouteqr.
\end{align}
\ew

We will now split the \ac{bvp} above into three steps \cite{loggAutomatedSolutionDifferential2012,chorinNumericalSolutionNavierStokes1968}:
\begin{enumerate}
        \titleditem{Approximated velocity}{
                \label{item_split_1}
                We introduce the velocity $\bvbarf$, which approximates the true velocity $\bvfr{}$, and which  satisfies the \ac{bvp} given by \cref{eq_fl_rigid_fl_1_reference_disc_aux} and \acp{bc}
                \bw
                \begin{align}
                        \label{bc_fl_rigid_1_reference_disc_aux}
                        \vbarf^1                                                     & = \vfin \; \text{ on }\pomineqr,                                                                                                         \\
                        \label{bc_fl_rigid_2_reference_disc_aux}
                        \vbarf^2                                                     & =  0 \; \text{ on } \pomineqr ,                                                                                                          \\
                        \label{bc_fl_rigid_3_reference_disc_aux}
                        \bvbarf                                                      & = {\bm 0} \; \text{ on } \pomweqr ,                                                                                                      \\
                        \label{bc_fl_rigid_4_reference_disc_aux}
                        \bvbarf                                                      & =
                        \omega^n  \left. \der{{\bm R}}{\theta} \right|_{\theta = \theta_n} \cdot (\bxr - \bfocalpoint)  \text{ on } \pomellipseeqr ,                                                                            \\
                        \label{bc_fl_rigid_5_reference_disc_aux}
                        \etaf \, G_{\beta 1}(\budom^{n-1}) \pderr{\Vf^\alpha}{\beta} & = 0                                                                                                           \; \text{ on } \pomouteqr.
                \end{align}
                \ew
                where $\bVf$ and $\sigmafr{\ast}$ are given by  \cref{eq_def_Vf,eq_def_sigmarast}.

        }
        \titleditem{Pressure correction}{
                \label{item_split_2}

                Subtracting \cref{eq_fl_rigid_fl_1_reference_disc,eq_fl_rigid_fl_1_reference_disc_aux} and neglecting $\odeltat$, we obtain \cref{eq_phi_1},        where the surface-tension increment $\phif$ is defined by \cref{eq_def_phi_pressure}.
                By applying $G_{\gamma \alpha}(\budom^{n-1}) \partial / \partial \xr[\gamma]$ to both sides of \cref{eq_phi_1} and using \cref{eq_fl_rigid_fl_2_reference_disc}, we obtain the second-order \ac{pde} \crefs{eq_poiss_like_phi} for $\phif$.

                We will now work out the \acp{bc} for \cref{eq_poiss_like_phi}. \Cref{bc_fl_rigid_6_reference_disc,eq_def_sigmarast,eq_def_phi_pressure} imply the \ac{bc}
                \be
                \label{bc_phi_1}
                \phif = 0 \, \text{ on } \pomouteqr.
                \ee
                We multiply \cref{eq_phi_1} by $G_{\gamma \alpha}(\budom^{n-1}) \normalr{\gamma}$, where $\normalr{}$ is the unit boundary normal in the \ac{ref} configuration pointing outside $\omr$, see \cref{fig_ref_cur_mem}A.  We obtain
                \begin{align}
                        \label{eq_phi_2}
                        \frac{\rhof}{\deltat} G_{\gamma \alpha}(\budom^{n-1}) \normalr{\gamma} (\vbarf^\alpha - \vfr{n, \, \alpha}) & = \newlinenn
                        \normalr{\gamma} G_{\gamma \alpha}(\budom^{n-1})  G_{\beta \alpha}(\budom^{n-1})  \,          \pderr{ \phif   }{\beta }.
                \end{align}
                Combining \cref{eq_phi_2} with \cref{bc_fl_rigid_1_reference_disc,bc_fl_rigid_2_reference_disc,bc_fl_rigid_3_reference_disc,bc_fl_rigid_4_reference_disc,bc_fl_rigid_1_reference_disc_aux,bc_fl_rigid_2_reference_disc_aux,bc_fl_rigid_3_reference_disc_aux,bc_fl_rigid_4_reference_disc_aux}, we obtain the second \ac{bc}
                \begin{align}
                        \label{bc_phi_2}
                        \normalr{\gamma}  G_{\gamma \alpha}(\budom^{n-1})  G_{\beta \alpha}(\budom^{n-1}) \,          \pderr{ \phif   }{\beta } & = \newlinenn 0
                        \text{ on } \pomineqr \cup \pomweqr \cup \pomellipseeqr                                                                 & .
                \end{align}

                The \ac{bvp} \crefs{eq_poiss_like_phi,bc_phi_1,bc_phi_2} determines $\phif$ and, through the definition \crefs{eq_def_phi_pressure}, the surface tension $\sigmafr{\nmhalf}$.
        }
        \titleditem{Velocity correction}{
                \label{item_split_3}
                The velocity $\bvfr{n}$ is determined from the approximate velocity $\bvbarf$ from \cref{eq_phi_1}.

        }
\end{enumerate}

\subsubsection{\Acl{dom}}\label{sec_splitting_scheme_fl_rig_dom}

The discrete version of  the \ac{bvp} for $\budom$, \cref{eq_mesh_fluid_rigid,bc_ela_1,bc_ela_2}, is given by \cref{eq_domain_fluid_rigid_disc} and
\begin{align}
        \label{bc_ela_1_disc}
        \budom^n & =\bzero \text{ on } \pomsqeqr,                                                                   \\
        \label{bc_ela_2_disc}
        \budom^n & = \bfocalpoint - \bxr + {\bm R}(\theta^n)\cdot (\bxr - \bfocalpoint) \text{ on } \pomellipseeqr.
\end{align}
The discrete version of the \ac{bvp} for $\budomdot$, \cref{eq_mesh_fluid_rigid_dot,bc_ela_1_dot,bc_ela_2_dot}, is given by \cref{eq_domain_fluid_rigid_disc_dot} and
\begin{align}
        \label{bc_ela_1_disc_dot}
        \budomdot  ^n & =\bzero \text{ on } \pomsqeqr,                                                         \\
        \label{bc_ela_2_disc_dot}
        \budomdot  ^n & =  \omega \der{{\bm R}}{\theta}\cdot (\bxr - \bfocalpoint) \text{ on } \pomellipseeqr.
\end{align}\\

We can now iterate in time for the ensemble of fields  of the fluid-body system as follows.
At each step $n$, given $\theta^{n-1}$, $\omega^{n-1}$, $\bvfr{n-1}$, $\sigmafr{\nmthalf}$, $\budom^{n-1}$ and $\budomdot^{n-1}$ from the preceding step, we
\begin{enumerate}
        \titleditem{Update the rigid body}{ \label{fl_rigid_step_1}
                Solve the algebraic equations \cref{eq_def_omega_disc,eq_fl_rigid_rigid_reference_disc} for $\theta^n$ and $\omega^n$.
        }
        \titleditem{Update the fluid domain}{ \label{fl_rigid_step_2}
                Solve  the \acp{bvp} \crefs{eq_domain_fluid_rigid_disc,bc_ela_1_disc,bc_ela_2_disc,eq_domain_fluid_rigid_disc_dot,bc_ela_1_disc_dot,bc_ela_2_disc_dot} for $\budom^n$ and $\budomdot^n$.
        }
        \titleditem{Update the fluid}{
                \label{fl_rigid_step_3}
                Solve the \acp{bvp} \crefs{eq_fl_rigid_fl_1_reference_disc_aux,bc_fl_rigid_1_reference_disc_aux,bc_fl_rigid_2_reference_disc_aux,bc_fl_rigid_3_reference_disc_aux,bc_fl_rigid_4_reference_disc_aux,bc_fl_rigid_5_reference_disc_aux} and \crefs{eq_phi_2,bc_phi_1,bc_phi_2}  for $\bvbarf$ and $\sigmafr{\nmhalf}$, and obtain $\bvfr{n}$ from \cref{eq_phi_1}.
        }
\end{enumerate}

The fields $\bvfc{n-1}$, $\sigmafc{\nmthalf}$ in the \ac{cur} configuration are then obtained from the respective fields in the \ac{ref} configuration through \cref{eq_vf_ref,eq_sigmaf_ref}.

\subsection{Characteristic time scale}
\label{sec_fluid_rigid_timescale}

\begin{figure}
        \centering
        \includegraphics[width=\columnwidth]{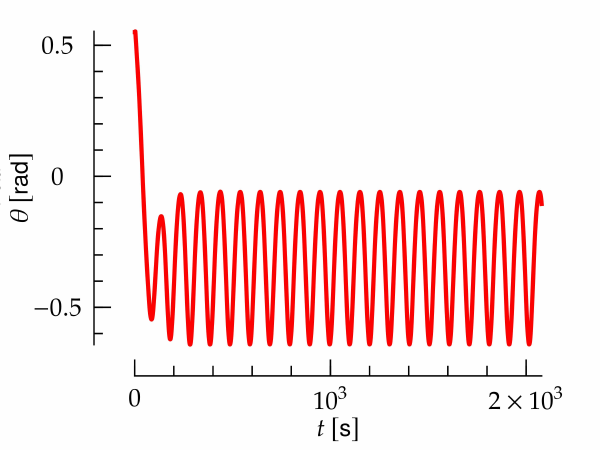}
        \caption{
                \label{fig_fluid_rigid_timescale}
                Angle of the \acl{bod} as a function of time for the example of \cref{fig_fluid_rigid}.
        }
\end{figure}

As shown in \cref{fig_fluid_rigid_timescale}, the dynamics of \cref{fig_fluid_rigid} involves a characteristic time scales $\tbody$, i.e., the period of the angular oscillations of the  body. This time scale can be estimated by considering the dynamical equation \crefs{eq_fl_rigid_rigid_current_1}, and approximating its \ac{lhs} as $I / \tbody^2$ and its \ac{rhs} as $ (\etaf v_0 /a \, 2 \pi a) a$. Here,  $\etaf v_0 /a$ is the viscous component of the force per unit length exerted by the fluid on the  body, cf. \cref{eq_strees_tensor_fluid}, $2 \pi a$ estimates the total length of the boundary of the  body, and $a$ approximates the lever arm of the force above. Putting everything together, we obtain
\be
\label{eq_expressions_tau_short}
\tbody \sim \sqrt{\frac{I}{2 \pi a \etaf v_0 }}.
\ee
For the parameters of \cref{fig_fluid_rigid}, \cref{eq_expressions_tau_short} yields $\tbody \sim 130 \, \second$, which is in agreement with the  period of the oscillations in \cref{fig_fluid_rigid_timescale}.

\section{Fluid and elastic body}\label{sec_fl_elastic_app}

\subsection{Equations of motion}
\label{app_sec_fl_elastic_equations_of_motion}

The equations of motion for the dynamics of the fluid and elastic body are the following.

First, the dynamics of the fluid is governed by the  \ac{ns} and mass-conservation equations \crefs{\eqsbfluc}, respectively. Their \acp{bc} are \cref{bc_fl_rigid_1_current,bc_fl_rigid_2_current,bc_fl_rigid_3_current,bc_fl_rigid_5_current,bc_fl_rigid_6_current}, which stay unchanged with respect to \cref{sec_fl_rigid}. On the other hand, the \ac{bc} \crefs{bc_fl_rigid_4_current}, which ensures that the fluid velocity matches the body velocity at the fluid-body interface, now reads
\be
\label{bc_fl_ela_4_current}
\bvfc{}(\bxc) = \left. \der{\buela(\bxr)}{t} \right|_{\bxr = \psi^t(\bxc)} \text{ on }  \pomcircouteqc
\ee

Second,  the motion of the  body is governed by Newton's equations of motion for an elastic medium \cite{landauTheoryElasticity1986}
\be\label{eq_fl_elastic_ela_current}
\rhoela \dern{\uela^\alpha(\bxr)}{t}{2} = \pderr{\Sela_{\alpha \beta}(\buela)}{\beta},
\ee
where  $\rhoela$ is the body density, which is assumed to be independent of space.
Also, $\buela$ is the body deformation field, which is defined along the lines of \cref{eq_def_u}.
The  body stress tensor is derived from a compressible neo-Hookean model \cite{bazilevsIsogeometricFluidstructureInteraction2008}, and it reads
\begin{align}
        \label{eq_def_S_neo_hookean}
        \Sela_{\alpha \beta} \equiv \frac{\muel}{\det F }\left(- \frac{1}{2} C_{\gamma \gamma} G_{\alpha \beta} + F_{\beta \alpha}\right) + & \newlinenn
        \kel [(\det(F))^2-1] G_{\alpha \beta},
\end{align}
where
\be
\label{eq_def_C}
C_{\alpha \beta } \equiv \delta_{\alpha \beta} + 2 E_{\alpha \beta}.
\ee
We have chosen this model because it is stable under compression, i.e., its potential energy diverges as $\det F \rightarrow 0$ \cite{bazilevsIsogeometricFluidstructureInteraction2008}. As opposed to linear models \cite{landauTheoryElasticity1986} and to the elastic model of  \cref{sec_eq_fluid_domain}, this model  yields a stable dynamics for the body as this is compressed  by  the fluid, see below and \cref{fig_fluid_ela}.
The \acp{bc} for \cref{eq_fl_elastic_ela_current} are
\begin{align}
        \label{bc_fl_ela_1_current}
        \buela(\bxr)                                                                                                              & =0               \text{ on }\pomcircineqr,   \\
        \label{bc_fl_ela_2_current}
        \left. \Sela_{\alpha \beta}(\buela^t)\right|_{\bxr = \bxi{s}} \epsilon_{\beta \gamma} \der{\xi^\gamma}{s}\                & =    \newlinenn
        \sigmastressc{\alpha}{\beta}(\bm{\varphi}^t(\bxi{s})) \epsilon_{\beta \gamma} \der{{\varphi}^{t, \, \gamma} (\bxi{s})}{s} & \text{ on } \pomcircouteqr.                &
\end{align}
In the \ac{lhs} of \cref{bc_fl_ela_2_current}, the quantity $\epsilon_{\beta \gamma} \der{\xi^\gamma}{s}$ represents the unit normal to $\pomcircouteqr$, and similarly for the \ac{rhs}.

Finally, the equations of motion for the fluid domain are \crefs{\eqsdom}, with \acp{bc} \crefs{bc_ela_1,bc_ela_1_dot} and
\begin{align}
        \label{bc_ela_3}
        \budom^t(\bxr)    & = \buela^t(\bxr) \text{ on } \pomcircouteqr,    \\
        \label{bc_ela_3_dot}
        \budomdot^t(\bxr) & = \bueladot^t(\bxr) \text{ on } \pomcircouteqr.
\end{align}
\Cref{bc_ela_3} and \cref{bc_ela_3_dot} ensure that the displacement and velocity of the  body and  fluid domain are conforming at the body-domain interface, and they replace  \cref{bc_ela_2,bc_ela_2_dot}, respectively.

\subsection{Dynamics}
\label{app_sec_fl_elastic_dynamics}

The system dynamics is obtained as follows: given $\bvfr{n-1}$, $\sigmafr{\nmthalf}$, $\buela^{n-1}$, $\bueladot^{n-1}$, $\budomdot^{n-1}$ and $\budomdot^{n-1}$  from the preceding step, we
\begin{itemize}
        \titleditem{Update the elastic body}{ \label{fl_ela_step_1}
                Solve for $\buela^n$ and $\bueladot^n$ the discrete version of \cref{eq_fl_elastic_ela_current}
                \begin{align}
                        \label{eq_fl_elastic_ela_current_disc_1}
                        \rhoela \frac{\ueladot^{n, \, \alpha} -  \ueladot^{n-1, \, \alpha}}{\deltat} & = \pderr{\Sela_{\alpha \beta}(\buela^n)}{\beta}, \\
                        \label{eq_fl_elastic_ela_current_disc_2}
                        \frac{\buela^n -  \buela^{n-1}}{\deltat}                                     & = \bueladot,
                \end{align}
                with \acp{bc} obtained from \cref{bc_fl_ela_1_current,bc_fl_ela_2_current}:
                \bw
                \begin{align}
                        \label{bc_fl_ela_1_current_disc}
                        \buela^n                                                                                                                                                                                                                & =  0          \text{ on }\pomcircineqr, \\
                        \label{bc_fl_ela_2_current_disc}
                        \epsilon_{\beta \gamma} \, \left. \Sela_{\alpha \beta}(\buela^n)\right|_{\bxr = \bxi{s}}\der{\xi^\gamma}{s}\                                                                                                            & =  \newlinenn
                        \sigmastressr{\alpha}{\beta}(\bxi{s};\bvfr{n-1}, \sigmafr{\nmthalf}, \buela^n) \epsilon_{\beta \gamma} \left. F_{\gamma \delta}(\bu_\ela^{n-1})\right|_{\bxr = \bxi{s}}\der{ \xi^\delta}{s}  \text{ on } \pomcircouteqr & .
                \end{align}
                \ew
                In \cref{bc_fl_ela_2_current_disc}, we have used \cref{eq_def_phi,eq_def_F} to rewrite the derivative of ${\bm \phi}(\bxi{s})$.

        }
        \titleditem{Update the fluid domain}{ \label{fl_ela_step_2}
                Solve for $\budom^n$ and $\budomdot^n$ the \acp{bvp} given by \cref{eq_domain_fluid_rigid_disc,eq_domain_fluid_rigid_disc_dot}  with \acp{bc} \crefs{bc_ela_1_disc}, \crefs{bc_ela_1_disc_dot} and
                \begin{align}
                        \label{bc_ela_3_disc}
                        \budom^n    & = \buela^n \text{ on } \pomcircouteqr,    \\
                        \label{bc_ela_3_dot_disc}
                        \budomdot^n & = \bueladot^n \text{ on } \pomcircouteqr.
                \end{align}
        }

        \titleditem{Update the fluid}{
                \label{fl_ela_step_3}
                Solve the \acp{bvp} given by  \crefs{eq_fl_rigid_fl_1_reference_disc_aux,eq_poiss_like_phi} and \acp{bc}
                \crefs{bc_fl_rigid_1_reference_disc_aux,bc_fl_rigid_2_reference_disc_aux,bc_fl_rigid_3_reference_disc_aux,bc_fl_rigid_5_reference_disc_aux,bc_phi_1,bc_phi_2} and
                \be
                \label{bc_fl_ela_4_current_disc}
                \bvbarf = \bueladot^n  \text{ on }  \pomcircouteqr,
                \ee
                which replaces \cref{bc_fl_rigid_5_reference_disc_aux} for a rigid body, and solve \cref{eq_phi_1}.

        }\end{itemize}

\begin{figure*}[t]
        \centering
        \includegraphics[width=\textwidth]{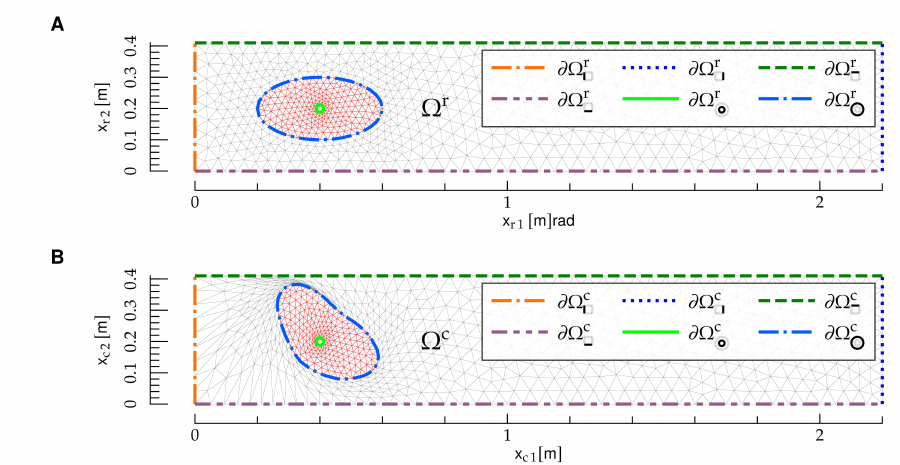}
        \caption{
                \label{fig_ref_curr_ela}
                \Acl{ref} and \acl{cur} configuration for the interaction between a  \acl{flu} and an \acl{ela}. The Figure follows the same notation as \cref{fig_ref_curr_rigid}.
        }
\end{figure*}

\bibliography{bibliography.bib}

@article{al-izziShearDrivenInstabilitiesMembrane2020,
  title = {Shear-{{Driven Instabilities}} of {{Membrane Tubes}} and {{Dynamin-Induced Scission}}},
  author = {{Al-Izzi}, Sami C. and Sens, Pierre and Turner, Matthew S.},
  year = 2020,
  month = jul,
  journal = {Phys. Rev. Lett.},
  volume = {125},
  number = {1},
  pages = {018101},
  issn = {0031-9007, 1079-7114},
  doi = {10.1103/PhysRevLett.125.018101},
  urldate = {2025-04-14},
  langid = {english}
}

@article{arroyoRelaxationDynamicsFluid2009i,
  title = {Relaxation Dynamics of Fluid Membranes},
  author = {Arroyo, M. and DeSimone, A.},
  year = 2009,
  month = mar,
  journal = {Phys. Rev. E Stat. Phys. Plasmas Fluids Relat. Interdiscip. Top.},
  volume = {79},
  number = {3},
  pages = {031915},
  issn = {1539-3755, 1550-2376},
  doi = {10.1103/PhysRevE.79.031915},
  urldate = {2024-09-11},
  langid = {english}
}

@article{barakatStokesFlowVesicles2018,
  title = {Stokes Flow of Vesicles in a Circular Tube},
  author = {Barakat, Joseph M. and Shaqfeh, Eric S. G.},
  year = 2018,
  month = sep,
  journal = {J. Fluid Mech.},
  volume = {851},
  pages = {606--635},
  issn = {0022-1120, 1469-7645},
  doi = {10.1017/jfm.2018.533},
  urldate = {2026-02-13},
  copyright = {https://www.cambridge.org/core/terms},
  langid = {english}
}

@article{barrettFiniteElementApproximation2016,
  title = {Finite Element Approximation for the Dynamics of Asymmetric Fluidic Biomembranes},
  author = {Barrett, John W. and Garcke, Harald and N{\"u}rnberg, Robert},
  year = 2016,
  month = aug,
  journal = {Math. Comput.},
  volume = {86},
  number = {305},
  pages = {1037--1069},
  issn = {0025-5718, 1088-6842},
  doi = {10.1090/mcom/3162},
  urldate = {2026-02-14},
  copyright = {https://www.ams.org/publications/copyright-and-permissions},
  langid = {english}
}

@article{Bassereau_2018,
  title = {The 2018 Biomembrane Curvature and Remodeling Roadmap},
  author = {Bassereau, Patricia and Jin, Rui and Baumgart, Tobias and Deserno, Markus and Dimova, Rumiana and Frolov, Vadim A and Bashkirov, Pavel V and Grubm{\"u}ller, Helmut and Jahn, Reinhard and Risselada, H Jelger and Johannes, Ludger and Kozlov, Michael M and Lipowsky, Reinhard and Pucadyil, Thomas J and Zeno, Wade F and Stachowiak, Jeanne C and Stamou, Dimitrios and Breuer, Art{\`u} and Lauritsen, Line and Simon, Camille and Sykes, C{\'e}cile and Voth, Gregory A and Weikl, Thomas R},
  year = 2018,
  month = jul,
  journal = {J. Phys. Appl. Phys.},
  volume = {51},
  number = {34},
  pages = {343001},
  publisher = {IOP Publishing},
  doi = {10.1088/1361-6463/aacb98}
}

@article{bazilevsIsogeometricFluidstructureInteraction2008,
  title = {Isogeometric Fluid-Structure Interaction: Theory, Algorithms, and Computations},
  shorttitle = {Isogeometric Fluid-Structure Interaction},
  author = {Bazilevs, Y. and Calo, V. M. and Hughes, T. J. R. and Zhang, Y.},
  year = 2008,
  month = dec,
  journal = {Comput. Mech.},
  volume = {43},
  number = {1},
  pages = {3--37},
  publisher = {{Springer Science and Business Media LLC}},
  issn = {0178-7675, 1432-0924},
  doi = {10.1007/s00466-008-0315-x},
  urldate = {2025-07-27},
  copyright = {http://www.springer.com/tdm},
  langid = {english}
}

@misc{blumFEAT2DFiniteElement1995,
  title = {{{FEAT2D}}, {{Finite Element Analysis Tools User Manual}}},
  author = {Blum, H. and Harig., J. and M{\"u}ller, S. and Schreiber, S. and Turek, S.},
  year = 1995
}

@article{boedecIsogeometricFEMBEMSimulations2017,
  title = {Isogeometric {{FEM-BEM}} Simulations of Drop, Capsule and Vesicle Dynamics in {{Stokes}} Flow},
  author = {Boedec, Gwenn and Leonetti, Marc and Jaeger, Marc},
  year = 2017,
  month = aug,
  journal = {J. Comput. Phys.},
  volume = {342},
  pages = {117--138},
  issn = {00219991},
  doi = {10.1016/j.jcp.2017.04.024},
  urldate = {2026-02-13},
  langid = {english}
}

@article{boffi2004stability,
  title = {Stability and Geometric Conservation Laws for {{ALE}} Formulations},
  author = {Boffi, Daniele and Gastaldi, Lucia},
  year = 2004,
  journal = {Comput. Methods Appl. Mech. Eng.},
  volume = {193},
  number = {42},
  pages = {4717--4739},
  issn = {0045-7825},
  doi = {10.1016/j.cma.2004.02.020}
}

@article{bonitoDynamicsBiomembranesEffect2011,
  title = {Dynamics of {{Biomembranes}}: {{Effect}} of the {{Bulk Fluid}}},
  shorttitle = {Dynamics of {{Biomembranes}}},
  author = {Bonito, A. and Nochetto, R.H. and Pauletti, M.S.},
  year = 2011,
  journal = {Math. Model. Nat. Phenom.},
  volume = {6},
  number = {5},
  pages = {25--43},
  issn = {0973-5348, 1760-6101},
  doi = {10.1051/mmnp/20116502},
  urldate = {2026-02-13},
  langid = {english}
}

@article{bonitoParametricFEMGeometric2010,
  title = {Parametric {{FEM}} for Geometric Biomembranes},
  author = {Bonito, Andrea and Nochetto, Ricardo H. and Sebastian Pauletti, M.},
  year = 2010,
  month = may,
  journal = {J. Comput. Phys.},
  volume = {229},
  number = {9},
  pages = {3171--3188},
  issn = {00219991},
  doi = {10.1016/j.jcp.2009.12.036},
  urldate = {2026-02-13},
  langid = {english}
}

@article{caiSoftPolydimethylsiloxaneElastomers2015,
  title = {Soft {{Poly}}(Dimethylsiloxane) {{Elastomers}} from {{Architecture}}-{{Driven Entanglement Free Design}}},
  author = {Cai, Li-Heng and Kodger, Thomas E. and Guerra, Rodrigo E. and Pegoraro, Adrian F. and Rubinstein, Michael and Weitz, David A.},
  year = 2015,
  month = sep,
  journal = {Adv. Mater.},
  volume = {27},
  number = {35},
  pages = {5132--5140},
  issn = {0935-9648, 1521-4095},
  doi = {10.1002/adma.201502771},
  urldate = {2026-01-27},
  copyright = {http://onlinelibrary.wiley.com/termsAndConditions\#vor},
  langid = {english}
}

@book{carmoDifferentialGeometryCurves1976,
  title = {{Differential geometry of curves and surfaces}},
  author = {do Carmo, Manfredo Perdig{\~a}o},
  year = 1976,
  publisher = {Prentice-Hall},
  address = {Englewood Cliffs, N.J},
  isbn = {978-0-13-212589-5},
  langid = {eng por},
  lccn = {QA641 .C33}
}

@article{causinAddedmassEffectDesign2005,
  title = {Added-Mass Effect in the Design of Partitioned Algorithms for Fluid--Structure Problems},
  author = {Causin, P. and Gerbeau, J.F. and Nobile, F.},
  year = 2005,
  month = oct,
  journal = {Comput. Methods Appl. Mech. Eng.},
  volume = {194},
  number = {42-44},
  pages = {4506--4527},
  issn = {00457825},
  doi = {10.1016/j.cma.2004.12.005},
  urldate = {2026-10-12},
  copyright = {https://www.elsevier.com/tdm/userlicense/1.0/},
  langid = {english}
}

@article{chippadaSimultaneousDeterminationYoungs2010,
  title = {Simultaneous Determination of {{Young}}'s Modulus, Shear Modulus, and {{Poisson}}'s Ratio of Soft Hydrogels},
  author = {Chippada, Uday and Yurke, Bernard and Langrana, Noshir A.},
  year = 2010,
  month = mar,
  journal = {J. Mater. Res.},
  volume = {25},
  number = {3},
  pages = {545--555},
  issn = {0884-2914, 2044-5326},
  doi = {10.1557/JMR.2010.0067},
  urldate = {2026-01-23},
  copyright = {https://www.cambridge.org/core/terms},
  langid = {english}
}

@article{chorinNumericalSolutionNavierStokes1968,
  title = {Numerical Solution of the {{Navier-Stokes}} Equations},
  author = {Chorin, Alexandre Joel},
  year = 1968,
  journal = {Math. Comput.},
  volume = {22},
  number = {104},
  pages = {745--762},
  issn = {0025-5718, 1088-6842},
  doi = {10.1090/S0025-5718-1968-0242392-2},
  urldate = {2025-04-14},
  langid = {english}
}

@article{crankPracticalMethodNumerical1947,
  title = {A Practical Method for Numerical Evaluation of Solutions of Partial Differential Equations of the Heat-Conduction Type},
  author = {Crank, J. and Nicolson, P.},
  year = 1947,
  month = jan,
  journal = {Math. Proc. Camb. Philos. Soc.},
  volume = {43},
  number = {1},
  pages = {50--67},
  issn = {0305-0041, 1469-8064},
  doi = {10.1017/S0305004100023197},
  urldate = {2025-05-15},
  copyright = {https://www.cambridge.org/core/terms},
  langid = {english}
}

@book{degennesCapillarityWettingPhenomena2004,
  title = {Capillarity and {{Wetting Phenomena}}},
  author = {De Gennes, Pierre-Gilles and {Brochard-Wyart}, Fran{\c c}oise and Qu{\'e}r{\'e}, David},
  year = 2004,
  publisher = {Springer New York},
  address = {New York, NY},
  doi = {10.1007/978-0-387-21656-0},
  urldate = {2026-01-15},
  copyright = {http://www.springer.com/tdm},
  isbn = {978-1-4419-1833-8 978-0-387-21656-0},
  langid = {english}
}

@article{derenyiFormationInteractionMembrane2002,
  title = {Formation and {{Interaction}} of {{Membrane Tubes}}},
  author = {Der{\'e}nyi, I. and J{\"u}licher, F. and Prost, J.},
  year = 2002,
  journal = {Phys. Rev. Lett.},
  volume = {88},
  number = {23},
  pages = {238101},
  issn = {0031-9007, 1079-7114},
  doi = {10.1103/PhysRevLett.88.238101},
  urldate = {2025-03-11},
  langid = {english}
}

@article{desernoFluidLipidMembranes2015,
  title = {Fluid Lipid Membranes: {{From}} Differential Geometry to Curvature Stresses},
  shorttitle = {Fluid Lipid Membranes},
  author = {Deserno, Markus},
  year = 2015,
  month = jan,
  journal = {Chem. Phys. Lipids},
  volume = {185},
  pages = {11--45},
  issn = {00093084},
  doi = {10.1016/j.chemphyslip.2014.05.001},
  urldate = {2025-06-14},
  langid = {english}
}

@misc{desernoNotesDifferentialGeometry2004,
  title = {Notes on {{Differential Geometry}}},
  author = {Deserno, M.},
  year = 2004,
  howpublished = {https://www.cmu.edu/biolphys/deserno/pdf/diff\_geom.pdf},
  langid = {english}
}

@article{diz-munozUseForceMembrane2013,
  title = {Use the Force: Membrane Tension as an Organizer of Cell Shape and Motility},
  shorttitle = {Use the Force},
  author = {{Diz-Mu{\~n}oz}, Alba and Fletcher, Daniel A. and Weiner, Orion D.},
  year = 2013,
  month = feb,
  journal = {Trends Cell Biol.},
  volume = {23},
  number = {2},
  pages = {47--53},
  issn = {09628924},
  doi = {10.1016/j.tcb.2012.09.006},
  urldate = {2026-11-12},
  copyright = {https://www.elsevier.com/tdm/userlicense/1.0/},
  langid = {english}
}

@incollection{doneaChapter14Arbitrary2004,
  title = {Chapter 14 {{Arbitrary Lagrangian}}--{{Eulerian Methods}}},
  booktitle = {The {{Encyclopedia}} of {{Computational Mechanics}}},
  author = {Donea, Jean and Huerta, Antonio and {Rodr{\i}guez-Ferran}, A},
  year = 2004,
  volume = {1},
  pages = {413--437},
  langid = {english}
}

@article{felippaPartitionedAnalysisCoupled2001,
  title = {Partitioned Analysis of Coupled Mechanical Systems},
  author = {Felippa, Carlos A. and Park, K.C. and Farhat, Charbel},
  year = 2001,
  month = mar,
  journal = {Comput. Methods Appl. Mech. Eng.},
  volume = {190},
  number = {24-25},
  pages = {3247--3270},
  issn = {00457825},
  doi = {10.1016/S0045-7825(00)00391-1},
  urldate = {2026-10-12},
  copyright = {https://www.elsevier.com/tdm/userlicense/1.0/},
  langid = {english}
}

@article{fengPreparationCharacterizationSilicone2017,
  title = {Preparation and Characterization of Silicone Rubber with High Modulus via Tension Spring-Type Crosslinking},
  author = {Feng, Linglong and Li, Shusheng and Feng, Shengyu},
  year = 2017,
  journal = {RSC Adv.},
  volume = {7},
  number = {22},
  pages = {13130--13137},
  issn = {2046-2069},
  doi = {10.1039/C7RA00293A},
  urldate = {2026-01-27},
  langid = {english}
}

@article{formaggia1999a,
  title = {A Stability Analysis for the {{Arbitrary Lagrangian Eulerian}} Formulation with Finite Elements},
  author = {Formaggia, Luca and Nobile, Fabio},
  year = 1999,
  journal = {East-West J. Numer. Math.},
  volume = {7},
  number = {2},
  pages = {105--131}
}

@article{forsterArtificialAddedMass2007,
  title = {Artificial Added Mass Instabilities in Sequential Staggered Coupling of Nonlinear Structures and Incompressible Viscous Flows},
  author = {F{\"o}rster, Christiane and Wall, Wolfgang A. and Ramm, Ekkehard},
  year = 2007,
  month = jan,
  journal = {Comput. Methods Appl. Mech. Eng.},
  volume = {196},
  number = {7},
  pages = {1278--1293},
  issn = {00457825},
  doi = {10.1016/j.cma.2006.09.002},
  urldate = {2026-10-12},
  copyright = {https://www.elsevier.com/tdm/userlicense/1.0/},
  langid = {english}
}

@article{godaMultistepTechniqueImplicit1979,
  title = {A Multistep Technique with Implicit Difference Schemes for Calculating Two- or Three-Dimensional Cavity Flows},
  author = {Goda, Katuhiko},
  year = 1979,
  month = jan,
  journal = {J. Comput. Phys.},
  volume = {30},
  number = {1},
  pages = {76--95},
  issn = {00219991},
  doi = {10.1016/0021-9991(79)90088-3},
  urldate = {2025-05-16},
  copyright = {https://www.elsevier.com/tdm/userlicense/1.0/},
  langid = {english}
}

@book{goldsteinClassicalMechanics2004,
  title = {Classical Mechanics},
  author = {Goldstein, H. and Poole, C. P. and Safko, J. L.},
  year = 2004,
  publisher = {Pearson}
}

@article{heilEfficientSolverFully2004,
  title = {An Efficient Solver for the Fully Coupled Solution of Large-Displacement Fluid--Structure Interaction Problems},
  author = {Heil, Matthias},
  year = 2004,
  month = jan,
  journal = {Comput. Methods Appl. Mech. Eng.},
  volume = {193},
  number = {1-2},
  pages = {1--23},
  issn = {00457825},
  doi = {10.1016/j.cma.2003.09.006},
  urldate = {2026-10-12},
  copyright = {https://www.elsevier.com/tdm/userlicense/1.0/},
  langid = {english}
}

@article{helfrichElasticPropertiesLipid1973,
  title = {Elastic {{Properties}} of {{Lipid Bilayers}}: {{Theory}} and {{Possible Experiments}}},
  shorttitle = {Elastic {{Properties}} of {{Lipid Bilayers}}},
  author = {Helfrich, W},
  year = 1973,
  month = dec,
  journal = {Z. Naturforschung C},
  volume = {28},
  number = {11-12},
  pages = {693--703},
  issn = {1865-7125, 0939-5075},
  doi = {10.1515/znc-1973-11-1209},
  urldate = {2025-04-21},
  langid = {english}
}

@book{hilsenrath1955tables,
  title = {Tables of Thermal Properties of Gases: Comprising Tables of Thermodynamic and Transport Properties of Air, Argon, Carbon Dioxide, Carbon Monoxide, Hydrogen, Nitrogen, Oxygen, and Steam},
  author = {Hilsenrath, Joseph},
  year = 1955,
  volume = {564},
  publisher = {US Department of Commerce, National Bureau of Standards}
}

@book{howellp.AppliedSolidMechanics2008,
  title = {Applied {{Solid Mechanics}}},
  author = {Howell, P. and Kozyreff, G. and Ockendon, J.},
  year = 2008,
  publisher = {Cambridge University Press}
}

@book{hsiungFirstCourseDifferential1981,
  title = {A {{First Course}} in {{Differential Geometry}}},
  author = {Hsiung, Chuan-Chih},
  year = 1981,
  publisher = {John Wiley \& Sons},
  address = {New York, USA},
  langid = {english}
}

@misc{kamenskyLectureNotesMAE,
  title = {{{Lecture notes for MAE 207: Finite element analysis for coupled problems}}},
  author = {Kamensky, David},
  year = 2022,
  langid = {english}
}

@article{kistlerCoherentExpandedAerogels1931,
  title = {Coherent {{Expanded Aerogels}} and {{Jellies}}},
  author = {KISTLER, S. S.},
  year = 1931,
  month = may,
  journal = {Nature},
  volume = {127},
  number = {3211},
  pages = {741--741},
  issn = {1476-4687},
  doi = {10.1038/127741a0}
}

@article{laadhariFiniteElementMethodSimulation2023,
  title = {Finite-{{Element Method}} for the {{Simulation}} of {{Lipid Vesicle}}/{{Fluid Interactions}} in a {{Quasi}}--{{Newtonian Fluid Flow}}},
  author = {Laadhari, Aymen},
  year = 2023,
  month = apr,
  journal = {Mathematics},
  volume = {11},
  number = {8},
  pages = {1950},
  issn = {2227-7390},
  doi = {10.3390/math11081950},
  urldate = {2026-02-13},
  langid = {english}
}

@book{lacarbonaraNonlinearStructuralMechanics2013,
  title = {Nonlinear {{Structural Mechanics}}: {{Theory}}, {{Dynamical Phenomena}} and {{Modeling}}},
  shorttitle = {Nonlinear {{Structural Mechanics}}},
  author = {Lacarbonara, Walter},
  year = 2013,
  publisher = {Springer US},
  address = {Boston, MA},
  doi = {10.1007/978-1-4419-1276-3},
  urldate = {2025-09-04},
  copyright = {https://www.springernature.com/gp/researchers/text-and-data-mining},
  isbn = {978-1-4419-1275-6 978-1-4419-1276-3},
  langid = {english}
}

@book{landauFluidMechanics1987,
  title = {Fluid {{Mechanics}}},
  author = {Landau, L. D. and Lifshitz, E. M.},
  year = 1987,
  publisher = {Pergamon}
}

@book{landauTheoryElasticity1986,
  title = {Theory of {{Elasticity}}},
  author = {Landau, L. D. and Lifshitz, {$E$}. M.},
  year = 1986,
  publisher = {Elsevier},
  langid = {english}
}

@article{lichtSynthetic3DPEGAnisogel2019,
  title = {Synthetic {{3D PEG-Anisogel Tailored}} with {{Fibronectin Fragments Induce Aligned Nerve Extension}}},
  author = {Licht, Christopher and Rose, Jonas C. and Anarkoli, Abdolrahman Omidinia and Blondel, Delphine and Roccio, Marta and Haraszti, Tam{\'a}s and Gehlen, David B. and Hubbell, Jeffrey A. and Lutolf, Matthias P. and De Laporte, Laura},
  year = 2019,
  month = nov,
  journal = {Biomacromolecules},
  volume = {20},
  number = {11},
  pages = {4075--4087},
  issn = {1525-7797, 1526-4602},
  doi = {10.1021/acs.biomac.9b00891},
  urldate = {2026-01-23},
  copyright = {http://pubs.acs.org/page/policy/authorchoice\_ccbyncnd\_termsofuse.html},
  langid = {english}
}

@book{loggAutomatedSolutionDifferential2012,
  title = {Automated {{Solution}} of {{Differential Equations}} by the {{Finite Element Method}}: {{The FEniCS Book}}},
  shorttitle = {Automated {{Solution}} of {{Differential Equations}} by the {{Finite Element Method}}},
  editor = {Logg, Anders and Mardal, Kent-Andre and Wells, Garth},
  year = 2012,
  series = {Lecture {{Notes}} in {{Computational Science}} and {{Engineering}}},
  volume = {84},
  publisher = {Springer Berlin Heidelberg},
  address = {Berlin, Heidelberg},
  doi = {10.1007/978-3-642-23099-8},
  urldate = {2025-06-19},
  copyright = {https://www.springernature.com/gp/researchers/text-and-data-mining},
  isbn = {978-3-642-23098-1 978-3-642-23099-8},
  langid = {english}
}

@book{malvernIntroductionMechanicsContinuous1969,
  title = {Introduction to the Mechanics of a Continuous Medium},
  author = {Malvern, L. E.},
  year = 1969,
  publisher = {Prentice Hall},
  address = {Englewood Cliffs, NJ}
}

@book{marchiafavaAppuntiDiGeometria2005,
  title = {{{Appunti di Geometria Differenziale}}},
  author = {Marchiafava, Stefano},
  year = 2005,
  volume = {I, II, III},
  publisher = {Edizioni Nuova Cultura}
}

@book{marsdenMathematicalFoundationsElasticity1994,
  title = {Mathematical Foundations of Elasticity},
  author = {Marsden, J. E. and Hughes, T. J. R.},
  year = 1994,
  publisher = {Dover},
  address = {New York},
  isbn = {978-0-486-67865-8},
  langid = {english},
  lccn = {QA931 .M42 1994}
}

@article{mullerQuickAccurateMethod2019,
  title = {A Quick and Accurate Method to Determine the {{Poisson}}'s Ratio and the Coefficient of Thermal Expansion of {{PDMS}}},
  author = {M{\"u}ller, Angelina and Wapler, Matthias C. and Wallrabe, Ulrike},
  year = 2019,
  journal = {Soft Matter},
  volume = {15},
  number = {4},
  pages = {779--784},
  issn = {1744-683X, 1744-6848},
  doi = {10.1039/C8SM02105H},
  urldate = {2026-01-27},
  langid = {english}
}

@article{niculescuUpdatedOverviewSilica2024,
  title = {An {{Updated Overview}} of {{Silica Aerogel-Based Nanomaterials}}},
  author = {Niculescu, Adelina-Gabriela and Tudorache, Dana-Ionela and Bocioag{\u a}, Maria and Mihaiescu, Dan Eduard and Hadibarata, Tony and Grumezescu, Alexandru Mihai},
  year = 2024,
  month = mar,
  journal = {Nanomaterials},
  volume = {14},
  number = {5},
  pages = {469},
  issn = {2079-4991},
  doi = {10.3390/nano14050469},
  urldate = {2026-01-23},
  langid = {english}
}

@article{pochivalovDevelopmentVibrationDamping2021,
  title = {Development of Vibration Damping Materials Based on Butyl Rubber: {{A}} Study of the Phase Equilibrium, Rheological, and Dynamic Properties of Compositions},
  shorttitle = {Development of Vibration Damping Materials Based on Butyl Rubber},
  author = {Pochivalov, Konstantin V. and Shilov, Alexander N. and Lebedeva, Tatyana N. and Ilyasova, Anna N. and Golovanov, Roman Yu. and Basko, Andrey V. and Kudryavtsev, Yaroslav V.},
  year = 2021,
  month = apr,
  journal = {J. Appl. Polym. Sci.},
  volume = {138},
  number = {15},
  pages = {50196},
  issn = {0021-8995, 1097-4628},
  doi = {10.1002/app.50196},
  urldate = {2026-01-27},
  langid = {english}
}

@incollection{Pugh_2016,
  title = {Soap Bubbles and Thin Films},
  booktitle = {Bubble and Foam Chemistry},
  author = {Pugh, Robert J.},
  year = 2016,
  pages = {84--111},
  publisher = {Cambridge University Press},
  address = {Cambridge}
}

@misc{reallGeneralRelativity,
  title = {General {{Relativity}}},
  author = {Reall, H.},
  year = 2022,
  howpublished = {https://www.damtp.cam.ac.uk/user/hsr1000/teaching.html}
}

@incollection{REHAGE1974161,
  title = {Elastic Properties of Crosslinked Polymers},
  booktitle = {Macromolecular Chemistry--9},
  author = {Rehage, G.},
  editor = {Cross, L.C.},
  year = 1974,
  pages = {161--178},
  publisher = {Butterworth-Heinemann},
  doi = {10.1016/B978-0-408-70654-4.50013-1},
  isbn = {978-0-408-70654-4}
}

@article{rodriguesSemiimplicitFiniteElement2015,
  title = {A Semi-Implicit Finite Element Method for Viscous Lipid Membranes},
  author = {Rodrigues, Diego S. and Ausas, Roberto F. and Mut, Fernando and Buscaglia, Gustavo C.},
  year = 2015,
  month = oct,
  journal = {J. Comput. Phys.},
  volume = {298},
  pages = {565--584},
  issn = {00219991},
  doi = {10.1016/j.jcp.2015.06.010},
  urldate = {2026-02-13},
  langid = {english}
}

@article{salbreuxActinCortexMechanics2012,
  title = {Actin Cortex Mechanics and Cellular Morphogenesis},
  author = {Salbreux, Guillaume and Charras, Guillaume and Paluch, Ewa},
  year = 2012,
  month = oct,
  journal = {Trends Cell Biol.},
  volume = {22},
  number = {10},
  pages = {536--545},
  issn = {09628924},
  doi = {10.1016/j.tcb.2012.07.001},
  urldate = {2025-06-15},
  copyright = {https://www.elsevier.com/tdm/userlicense/1.0/},
  langid = {english}
}

@article{sanmartinConvergenceFiniteElement2009,
  title = {Convergence of a Finite Element/{{ALE}} Method for the {{Stokes}} Equations in a Domain Depending on Time},
  author = {San Mart{\'i}n, Jorge and Smaranda, Loredana and Takahashi, Tak{\'e}o},
  year = 2009,
  month = aug,
  journal = {J. Comput. Appl. Math.},
  volume = {230},
  number = {2},
  pages = {521--545},
  issn = {03770427},
  doi = {10.1016/j.cam.2008.12.021},
  urldate = {2026-11-12},
  copyright = {https://www.elsevier.com/tdm/userlicense/1.0/},
  langid = {english}
}

@misc{schaferBenchmarkComputationsLaminar1996,
  title = {Benchmark {{Computations}} of {{Laminar Flow Around}} a {{Cylinder}}},
  author = {Schafer, M and Turek, S},
  year = 1996
}

@article{seifertConfigurationsFluidMembranes1997a,
  title = {Configurations of Fluid Membranes and Vesicles},
  author = {Seifert, Udo},
  year = 1997,
  month = feb,
  journal = {Adv. Phys.},
  volume = {46},
  number = {1},
  pages = {13--137},
  issn = {0001-8732, 1460-6976},
  doi = {10.1080/00018739700101488},
  urldate = {2026-11-12},
  langid = {english}
}

@article{sensMembraneTensionCytoskeleton2015,
  title = {Membrane Tension and Cytoskeleton Organization in Cell Motility},
  author = {Sens, Pierre and Plastino, Julie},
  year = 2015,
  month = jul,
  journal = {J. Phys. Condens. Matter},
  volume = {27},
  number = {27},
  pages = {273103},
  issn = {0953-8984, 1361-648X},
  doi = {10.1088/0953-8984/27/27/273103},
  urldate = {2026-11-12},
  langid = {english}
}

@article{simonActinDynamicsDrive2019,
  title = {Actin Dynamics Drive Cell-like Membrane Deformation},
  author = {Simon, Camille and Kusters, R{\'e}my and Caorsi, Valentina and Allard, Antoine and {Abou-Ghali}, Majdouline and Manzi, John and Di Cicco, Aur{\'e}lie and L{\'e}vy, Daniel and Lenz, Martin and Joanny, Jean-Fran{\c c}ois and Campillo, Cl{\'e}ment and Plastino, Julie and Sens, Pierre and Sykes, C{\'e}cile},
  year = 2019,
  month = jun,
  journal = {Nat. Phys.},
  volume = {15},
  number = {6},
  pages = {602--609},
  issn = {1745-2473, 1745-2481},
  doi = {10.1038/s41567-019-0464-1},
  urldate = {2025-06-15},
  langid = {english}
}

@book{slaughterLinearizedTheoryElasticity2002,
  title = {The {{Linearized Theory}} of {{Elasticity}}},
  author = {Slaughter, William S.},
  year = 2002,
  publisher = {Birkh\"auser Boston},
  address = {Boston, MA},
  doi = {10.1007/978-1-4612-0093-2},
  urldate = {2025-06-24},
  copyright = {http://www.springer.com/tdm},
  isbn = {978-1-4612-6608-2 978-1-4612-0093-2},
  langid = {english}
}

@book{SolvingOrdinaryDifferential1993,
  title = {Solving {{Ordinary Differential Equations I}}},
  author = {Hairer, E. and N{\o}rsett, S. P. and Wanner, G.},
  year = 1993,
  series = {Springer {{Series}} in {{Computational Mathematics}}},
  volume = {8},
  publisher = {Springer Berlin Heidelberg},
  address = {Berlin, Heidelberg},
  doi = {10.1007/978-3-540-78862-1},
  urldate = {2026-01-01},
  copyright = {http://www.springer.com/tdm},
  isbn = {978-3-540-56670-0 978-3-540-78862-1},
  langid = {english}
}

@article{vasanMechanicalModelReveals2020,
  title = {A Mechanical Model Reveals That Non-Axisymmetric Buckling Lowers the Energy Barrier Associated with Membrane Neck Constriction},
  author = {Vasan, R. and Rudraraju, S. and Akamatsu, M. and Garikipati, K. and Rangamani, P.},
  year = 2020,
  journal = {Soft Matter},
  volume = {16},
  number = {3},
  pages = {784--797},
  issn = {1744-683X, 1744-6848},
  doi = {10.1039/C9SM01494B},
  urldate = {2026-11-13},
  langid = {english}
}

@article{worthmullerIRENEFluIdLayeR2025,
  title = {{{IRENE}}: A {{fluId layeR finitE-elemeNt softwarE}}},
  shorttitle = {{{IRENE}}},
  author = {W{\"o}rthm{\"u}ller, Dennis and Ferraro, Gaetano and Sens, Pierre and Castellana, Michele},
  year = 2027,
  month = jan,
  journal = {Comput. Phys. Commun.},
  volume = {330},
  pages = {110366},
  issn = {00104655},
  doi = {10.1016/j.cpc.2026.110366},
  urldate = {2026-10-13},
  langid = {english}
}

@article{xuConcurrentStiffeningSoftening2023,
  title = {Concurrent Stiffening and Softening in Hydrogels under Dehydration},
  author = {Xu, Shuai and Zhou, Zidi and Liu, Zishun and Sharma, Pradeep},
  year = 2023,
  month = jan,
  journal = {Sci. Adv.},
  volume = {9},
  number = {1},
  pages = {eade3240},
  issn = {2375-2548},
  doi = {10.1126/sciadv.ade3240},
  urldate = {2026-01-23},
  langid = {english}
}

@article{zhong-canBendingEnergyVesicle1989,
  title = {Bending Energy of Vesicle Membranes: {{General}} Expressions for the First, Second, and Third Variation of the Shape Energy and Applications to Spheres and Cylinders},
  shorttitle = {Bending Energy of Vesicle Membranes},
  author = {{Zhong-can}, Ou-Yang and Helfrich, Wolfgang},
  year = 1989,
  month = may,
  journal = {Phys. Rev. A},
  volume = {39},
  number = {10},
  pages = {5280--5288},
  issn = {0556-2791},
  doi = {10.1103/PhysRevA.39.5280},
  urldate = {2025-04-22},
  copyright = {http://link.aps.org/licenses/aps-default-license},
  langid = {english}
}

@misc{zhouModelingSimulationOpen2026,
  title = {Modeling and {{Simulation}} of {{Open Membranes}} in {{Stokes Flow}} with {{Mixed-Dimensional Coupling}}},
  author = {Zhou, Han and Young, Yuan-Nan and Mori, Yoichiro},
  year = 2026,
  month = jan,
  number = {arXiv:2504.16823},
  eprint = {2504.16823},
  primaryclass = {math},
  publisher = {arXiv},
  doi = {10.48550/arXiv.2504.16823},
  urldate = {2026-02-14},
  archiveprefix = {arXiv},
  langid = {english}
}

\end{document}